\documentclass[useAMS,usenatbib,onecolumn]{mn2e}

\usepackage{mathtext,amssymb}
\usepackage{epsfig}
\usepackage{graphics}
\usepackage{amsmath}

\renewcommand{\vector}[1]{\ensuremath{\mathbf{#1}}}

\newcommand{\mdot}{\ensuremath{\dot{m}}}

\newcommand{\Msun}{\ensuremath{\rm M_\odot}}
\newcommand{\Msunyr}{\ensuremath{\rm M_\odot \, yr^{-1}}}
\newcommand{\Gyr}{\ensuremath{\rm Gyr}}

\begin{document}
\title[Black Hole Spin Evolution]{Black Hole Spin Evolution Affected by Magnetic Field Decay}
\author[]{Anna Chashkina \thanks{E-mail: sagitta.minor@gmail.com} and Pavel Abolmasov \\
Sternberg Astronomical Institute, Moscow State University,
  Moscow, Russia 119992\\}

\date{Accepted ---. Received ---; in
  original form --- }

\pagerange{\pageref{firstpage}--\pageref{lastpage}} \pubyear{2009}

\maketitle

\label{firstpage}

\begin{abstract}
Black holes are spun up by accreting matter and possibly  spun-down by magnetic fields. In our work we consider the effect on black hole rotation of the two electromagnetic processes, Blandford-Znajek and Direct Magnetic Link, that differ in their magnetic field configuration. The efficiency of these processes varies with mass accretion rate and accretion regime and  generally result in an equilibrium spin parameter in the range from $0.35$ to $\sim 0.98$. Magnetic field loses its energy while being accreted that may lead to an
 increase in equilibrium Kerr parameter for the case of advection-dominated
 disc.  We find magnetic field decay decay can decrease electromagnetic
 term significantly thus increasing the Kerr parameter. 
We have performed Monte-Carlo simulations for a supermassive black hole
population. Our simulations show broad distributions in Kerr parameter
($0.1\lesssim a \lesssim0.98$) with a peak at $a\sim 0.6$. To explain the high
observational Kerr parameter values of $a \gtrsim 0.9$, episodes of
supercritical accretion are required.
This implication does not however take into account black hole mergers (that play an important role for supermassive black hole evolution).
\end{abstract}


\section{Introduction}

Stellar mass black holes (BH) are the final stage of the evolution of the most massive stars. Supermassive black holes (SMBH) can be formed by accretion onto stellar mass BH or by
 mergers of stellar mass and intermediate mass BHs, but this question now is open (see \citet{shankar09} and \citet{haiman13} for review).
 Any BH is characterized by its mass $M$ and its total angular momentum $J$. These parameters may evolve under the influence of different processes, 
primarily due to accretion of matter. There are two types of objects where the black hole mass grows significantly during their evolution: black holes in some 
X-ray binaries such as X-ray novae and SMBH in galactic nuclei. 

Black holes are believed to be spun up by gas accretion. { For the standard thin disc case }the net angular
momentum of the material at the innermost stable circular Keplerian orbit (having radius $R_{in}$) always exceeds this of the black hole itself, the latter limited by the value $GM/c$. Below we will use relativistic dimensionless quantities $j^\dagger =L^\dagger(R_{in}) / (GM/c)$ (specific angular momentum in $GM/c$ units) and $E^\dagger$ (dimensionless specific energy at infinity for a particle at $R_{in}$) following the designations used by \citet{PT74}. We also describe BH rotation in terms of its Kerr parameter  $a = J c /GM^2 < 1$.

 Inside $R_{in}$ accreting matter conserves its angular momentum, hence the black hole is spun up  according to a simple law $da/d\ln M = \left(\displaystyle\frac{j^\dagger}{E^\dagger}-2a\right)
 \times  \mu$, where $\mu\simeq 1$ is dimensionless specific enthalpy \citep{BAN97} and accounts for
 internal energy contribution to the total (relativistic) mass
 accretion rate.
 In advection-dominated regime and supercritical accretion these
  assumptions are violated due to non-Keplerian rotation and transonic
  structure of the disc, 
where angular momentum can be transferred by stresses inside the last stable
orbit (see \citet{Abramowicz2010} and sec. 3).

Black hole spin evolution has been considered by more than forty years, since \citet{bardeen70} obtained an analytical solution for black hole spin-up by accretion. Since then, several processes were proposed that may stop this spin-up at higher or lower spin values.
First of all, selective capture of accretion disc radiation in Kerr metric is able to provide a strong counteracting spin-down torque if the BH rotation parameter is close to maximal. Due to this reason, it is impossible to spin
up black holes to very high $a$ \citep{thorne74}. Dimensionless spin
parameter stalls at a value $\sim 0.998$, when the spin-up is compensated by the angular
momentum of the captured photons emitted by the inner parts of the flow.

 The work by \citet{Kato09} proposes that
Blandford-Znajek process \citep{BZ77} may be responsible for black hole spin settling somewhere around $0.4-0.5$. The power of this process is determined by the Poynting flux generated by poloidal magnetic field and toroidal electric field induced by black hole rotation.  Radio-bright active galactic nuclei (AGNs) and microquasars are
supposed to power their jets via the Blandford-Znajek mechanism. 
At the end of XX century alternative mechanism known as direct magnetic link was proposed (see \citet{uzdensky05} and references therein).
 Below we use the name DML (direct magnetic link) for any process of magnetic-field mediated angular momentum exchange between the BH and accretion disc.

 Magnetic field in this case connects the black hole with the accretion disc. Depending on relation between the angular velocities of the black hole and the accretion disc, black hole can spin-up or spin-down.  This mechanism differs from Blandford-Znajek process in its magnetic field geometry (see the next section and \citet{uzdensky05}, \citet{krolik1999}).
If magnetic flux through the disc is not zero, flux can be accumulated inside the last stable orbit that leads to Blandford-Znajek geometry. This is confirmed by many simulations (see for example \citet{Tchekhovskoy12}). In our work we will consider black hole spin evolution taking into account both electromagnetic mechanisms: Blandford-Znajek process and DML.
 We will describe these processes in detail in the next section.

It should be noted that magnetic fields dissipate at the vicinity of the event horizon. Below, we use membrane paradigm \citep{membrane} to estimate ``Joule
losses'' at the stretched horizon and the power that magnetic fields extract
from black hole rotation. 
One actually needs accretion to supply the magnetic field. In this paper, we estimate the
effect of magnetic field decay for constant mass accretion rate. 
Note that magnetic field can not dissipate in BZ case because magnetic flux is
conserved and the field can be supported by currents in the disc.

In the following section, we discuss the basic properties of the basic electromagnetic field configurations, in section \ref{sec:rev} we use membrane paradigm to estimate the angular
momentum and energy extracted from a rotating black hole by magnetic
fields for these configurations. In section \ref{sec:decay}, we
discuss the influence of magnetic field decay. 
We present our results in section \ref{sec:results} and discuss them in section \ref{sec:discussion}.
{ In section \ref{sec:results} we also present some applications supermassive and stellar mass black holes.}

\section{Power and Spin-Down Rate}

In our work we use Membrane paradigm \citep{membrane} that helps describe the
black hole horizon avoiding complicated general relativity calculations.

Recent numerical results such a jet simulations support the capability of this approach to reproduce the main features of black hole magnetospheres and accretion flows

\citep{Narayan2013}. \\
In the Membrane approach framework \citep{membrane}, the event horizon is considered
 surrounded by the so-called stretched horizon -- viscous conducting sphere that has finite entropy and does not conduct heat. 
This sphere rotates with angular velocity $\Omega_H= \displaystyle\frac{a}{2}\displaystyle \frac{c}{R_H}$, here  $R_H =
\left( 1+ \sqrt{1-a^2}\right) GM/c^2=r_H GM/c^2$ is horizon radius. Using this paradigm, properties  of a black hole can be considered without complicated general relativistic calculations.
  This approach is sufficient for our needs.

There are two main electromagnetic processes that affect the rotational
evolution of a black hole: Blandford-Znajek  process \citep{BZ77} and direct
magnetic link \citep{uzdensky05}. The main difference  between these processes
is in magnetic field configuration. Blandford-Znajek process works when field
lines connect the black hole with a distant region such as  jet \footnote{
   We do not need the jet to be collimated for the Blandford-Znajek
    process to operate. Jet collimation occurs at the distances much larger
    than characteristic scales where black hole rotation energy is converted
    into electromagnetic \citep{Tchekhovskoy10}. Collimation can be produced
    by  some surrounding media such as thick disc or intense disk
    wind. Numerical simulations such as \citet{Fragile12} also suggest that
    jet collimation is not directly connected to disc thickness but rather to
    existence of disc corona or wind providing the necessary pressure. } 
(Fig. 1, right). In DML case magnetic lines connect the  stretched horizon
with the accretion disc (Fig. 1, left).

\begin{figure}\label{processes}
\includegraphics[width=0.5\columnwidth]{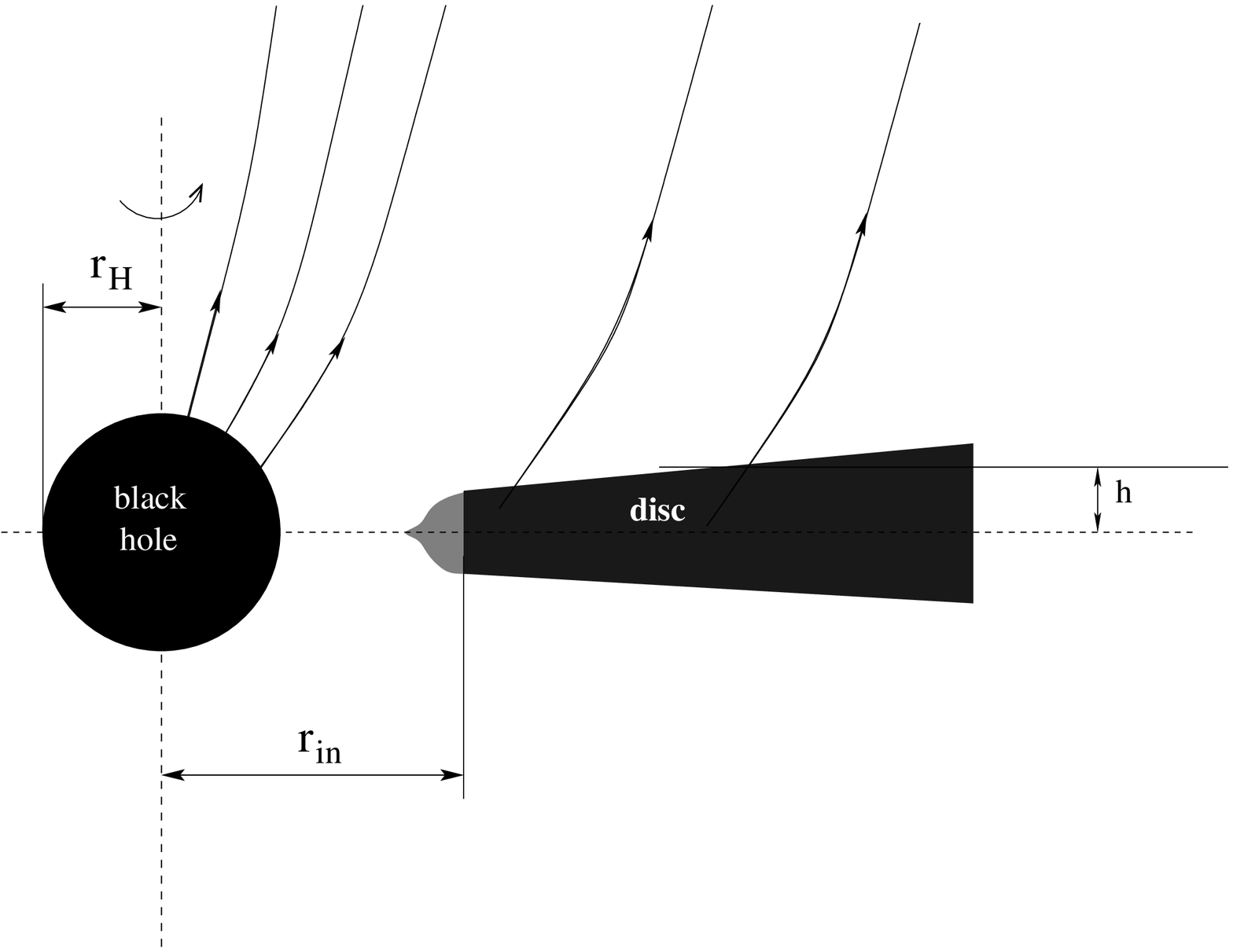}
\includegraphics[width=0.5\columnwidth]{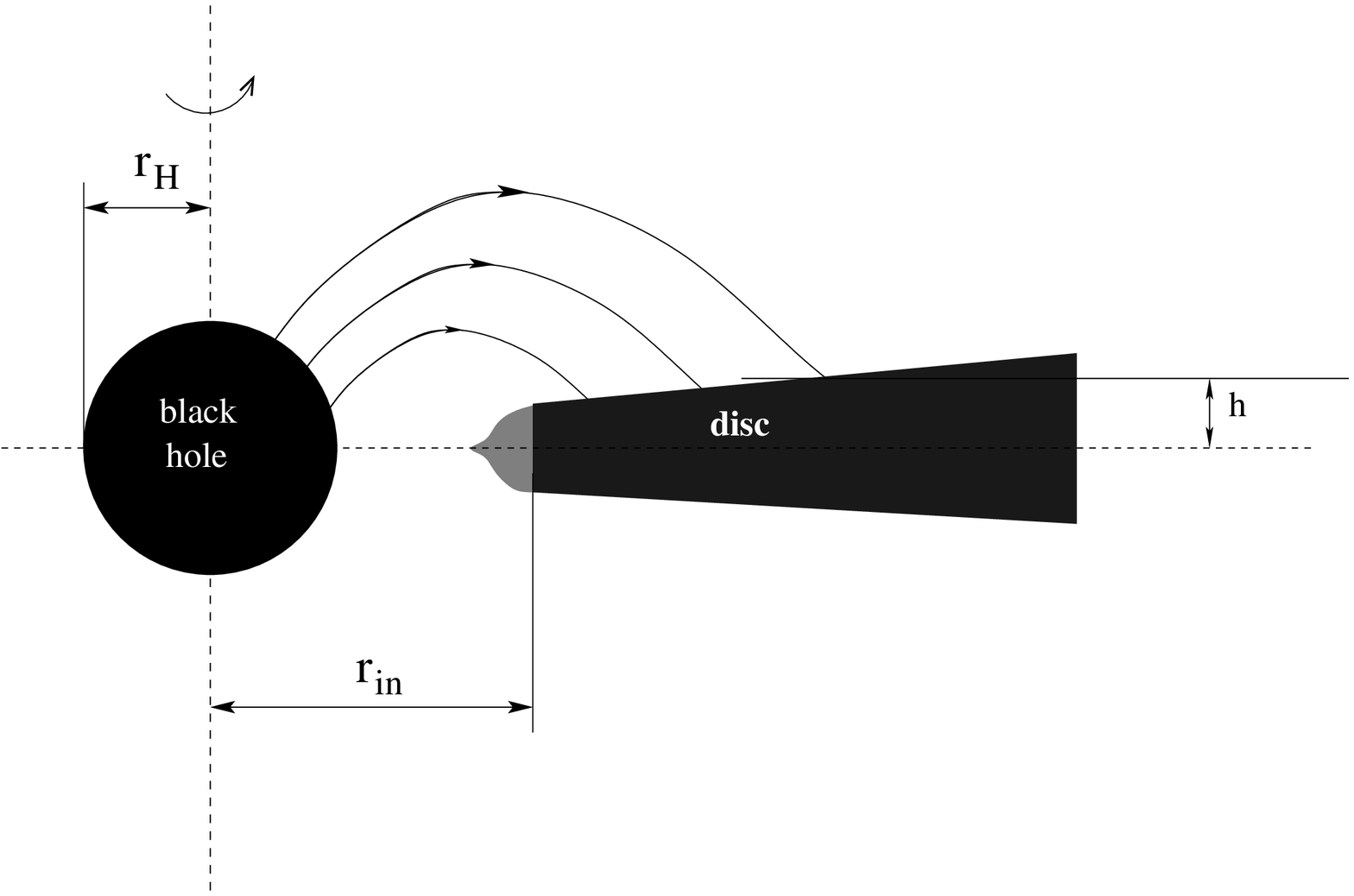}
\caption{Magnetic field configuration for Blandford-Znajek (left) and  Direct Magnetic Link (right) cases.}
\end{figure}

One can describe a black hole magnetosphere   as a steady and axisymmetric system of nested electric circuits consisting of a battery and a load. The battery here is the warped space region near the black hole, and the jet or accretion disc plays the role of the load.

Poloidal magnetic field: 

\begin{equation}
\vec{B_p}=\frac{({\nabla \Psi}) \times e_{\hat{\phi}}}{2 \pi \varpi}
\end{equation} 
here $\Psi$ is magnetic flux, $\varpi$ is the radial cylindric coordinate.
Field lines in such configuration rotate as a rigid body with Ferraro angular velocity $\Omega_F$ \citep{ferraro37}. In a force-free magnetosphere magnetic field is degenerate, $\vec{E}\cdot \vec{B}=0$, and one can write  electric field as follows:
\begin{equation}
\vec{E}=-\frac{(\Omega_F-\omega)}{2\pi\alpha_{l}}\nabla{\Psi}
\end{equation}
where $\omega$ is Lense-Thirring precession frequency and $\alpha_{l}$ is the lapse function taking in account dilation of time.

 The electromotive force produced by the gravitational field:\\

\begin{equation}
\Delta V=\oint \alpha_l \vec{E}d\vec{l}=\frac{1}{2\pi}\Omega_H \Delta \Psi
\end{equation}

This electromotive force is balanced by the potential difference along the horizon surface and potential difference in the load region:
\begin{equation} \label{eqq:all}
\Delta V=\Delta V_H+\Delta V_L
\end{equation}

The potential difference along the horizon part of the current:

\begin{equation}\label{eqq:horizon}
\Delta V_{H}=\oint \alpha_{l} \vec{E}d\vec{l}=\frac{1}{2\pi}(\Omega_H-\Omega_F )\Delta \Psi
\end{equation}

One can find a potential difference within the load using eq. (\ref{eqq:all}) and eq. (\ref{eqq:horizon}) as follows:
\begin{equation}
\Delta V_{L}=\frac{1}{2\pi}\Omega_F \Delta \Psi
\end{equation}
  
Resistance of a  horizon belt of latitudinal size $\Delta l$:
\begin{equation}
\Delta R_{H}=R_{H}\displaystyle\frac{\Delta l}{2\pi \varpi}=R_H\displaystyle\frac{\Delta \Psi}{4\pi^2\varpi^2 B_n}
\end{equation}
here $R_H=4\pi/c=377$~Ohm is specific horizon surface resistivity, $B_n$ is normal component of magnetic field at the horizon, $\Delta l$ is elementary distance along the horizon between two magnetic layers.
Since $I=$~const inside a given flux tube,
\begin{equation} 
\displaystyle\frac{\Omega_F}{\Omega_H-\Omega_F}=\displaystyle\frac{\Delta V_L}{\Delta V_H}=\displaystyle\frac{\Delta R_L}{\Delta R_H}
\end{equation}

Taking into account that $\Delta V_L=I\Delta R_L$ and  $\Delta V_H=I\Delta R_H$, one can write:
\begin{equation}
I=\displaystyle\frac{\Delta V}{\Delta R_H+\Delta R_L}=\displaystyle\frac{1}{2}(\Omega_H-\Omega_F)\varpi^2 B_n
\end{equation}

Total Joule losses for an elementary circuit are $\Delta P=I\Delta V$.

Dissipative power at the horizon:

\begin{equation}
\Delta P_H=\displaystyle\frac{(\Omega_H-\Omega_F)^2}{4\pi}\varpi^2 B_n\Delta \Psi
\end{equation}

Power transferred by Poynting vector towards the load equals:

\begin{equation}
\Delta P_L=\displaystyle\frac{\Omega_F(\Omega_H-\Omega_F)}{4\pi}\varpi^2 B_n\Delta \Psi
\end{equation}

The total energy losses towards the load may be found by integrating $\Delta P_L$ over the magnetosphere:

\begin{equation}\label{E:power:gen}
\frac{dE}{dt} = \int_{\Psi}\delta P_{L}= \displaystyle\frac{1}{8c} \Omega_F (\Omega_F - \Omega_H) r_H^4
\langle B_p^2 \rangle
\end{equation}

Angular momentum losses may be calculated by integrating the $r\varphi$- component of the Maxwell stress tensor:

\begin{equation}
\frac{dJ}{dt}=\int_{\Psi}\displaystyle\frac{1}{4\pi}B_{\phi}B_{p}\cdot 2\pi\varpi^{2}\delta\Psi=\displaystyle \frac{1}{8c} (\Omega_F - \Omega_H) r_H^4
\langle B_p^2 \rangle
\end{equation}

Since energy and angular momentum losses differ by the factor of $\Omega_{F}$, we can interpret them as extraction of rotational energy from the black hole.

Depending on the difference between $\Omega_F$ and $\Omega_H$, energy
and angular momentum may be either absorbed by the black hole or
extracted from it by magnetic fields. The main difference between the
two configurations, Blandford-Znajek and direct magnetic link,
is in the value of Ferraro frequency. For the
Blandford-Znajek case, relation between $\Omega_H$ and $\Omega_F$ is
determined by the unknown resistance of the distant load. Below 
we assume $\Omega_F=\Omega_H/2$ in the case of
Blandford-Znajek configuration, that corresponds to maximal energy
extraction rate from black hole rotation.

We assume that magnetic field is in equipartition with gas+radiation as $\displaystyle\frac{B^2_p}{8\pi}=\displaystyle\frac{1}{\beta}p$, where $p$  is total thermal pressure, and $\beta$ is a dimensionless coefficient about unity. \\
Finally, we write the two dynamical
equations (12,13) for Blandford-Znajek process as follows:

\begin{equation}\label{E:power}
\frac{dE}{dt} = \displaystyle\frac{\pi}{\beta}\displaystyle \frac{\Omega_H^2 R_H^4 p}{4c}
\end{equation}

\begin{equation}\label{E:spin}
\frac{dJ}{dt} =\displaystyle\frac{\pi}{\beta} \displaystyle\frac{\Omega_H R_H^4 p}{2c}
\end{equation}

In the case of direct magnetic link the effective black hole conductivity is much smaller
than the conductivity of the ionized plasma of the disc. Therefore,
magnetic field may be considered frozen into the disc at some effective
radius $r_{eff}$ close to the last stable orbit radius $r_{in}$. The pressure inside 
a standard disc (in gas pressure dominated zone, \citep{SS73} is close to the maximal value at the distance of $1.3r_{in}$, hence
we assume $r_{eff}=1.3r_{in}$ in our calculations. Ferraro frequency equals Kepler
frequency at the effective radius, $\Omega_F=\Omega_K(r_{eff})$. 

 For higher rotation
parameters, Keplerian frequency at the inner face of the disc is
lower than the frequency $\Omega_H$, and black hole rotation energy
is transferred to the disc. The innermost stable orbit
corotates with the stretched horizon at $a=a_{cr} \simeq
0.3594$ \citep{uzdensky05}, for $r_{eff}=1.3r_{in}$, the corresponding
critical value is $a_{cr} \sim 0.218$. It can be checked that for $a>a_{cr}$ magnetic field
slows down or stops black hole spin-up, and sometimes is proposed to
stop accretion \citep{agolkrolik} by creating a non-zero torque at the
inner boundary of the disc.

Magnetic field lines in the case of direct magnetic link connect disc to the stretched
horizon, therefore the magnetic field strength should be modified by a
factor determined by geometry. Magnetic flux conservation implies:

$$
R_H h_H B_p^{H} = 2 h R_{eff} B_p^{disc}
$$

Here, $h_H$ is the width of the equatorial band at the horizon surface
that is connected to the disc, $h$ is disc half-thickness at the $r_{eff}$. Left- and right-hand sides correspond
to the flux at the horizon and at the surface of the disc at $r_{eff}$. Finally,
the energy and angular momentum transmitted by direct magnetic link
may be expressed in a form similar to eqs. (\ref{E:power},\ref{E:spin}):

\begin{equation}\label{E:power:dml}
\frac{dE}{dt} = \displaystyle\frac{\pi}{\beta c} \Omega_K(r_{eff}) (\Omega_K(r_{eff}) - \Omega_H) R_H^2 R^2_{eff}p \left(\displaystyle\frac{h}{h_{H}}\right)^{2}
\end{equation}

\begin{equation}\label{E:spin:dml}
\frac{dJ}{dt} = \displaystyle\frac{\pi}{\beta c} (\Omega_K(r_{eff}) - \Omega_H) R_H^2 R^2_{eff}p\left(\displaystyle\frac{h}{h_{H}}\right)^{2}
\end{equation}

In our calculations we assume $h_H=R_H$.

\bigskip

{ The question may arise whether these two magnetic field configurations are the only ones allowed for accretion disks. The geometry of seed magnetic fields may be different but as long as we consider axisymmetric configurations such as accretion disks the possible zoo of magnetic field configurations is limited. One can expand magnetic field in multipoles. In general, every magnetic field configuration will consist of some uniform part and higher multipoles that should be confined inside the falling matter. Note that magnetic loop curvature leads to Lorentz force that makes some configurations unstable (such as the conventional DML case shown in figure 1, right panel). Our consideration of DML does not depend on the characteristic magnetic field spatial scales in the disk as long as radial motions are much faster inside the last stable orbit.}

\section{Rotational evolution}\label{sec:rev}

In this section we estimate the electromagnetic terms in rotational evolution
for different accretion regimes.  It is convenient to normalize all
characteristic timescales to Eddington time $t_{Edd} = c\varkappa/4\pi G \sim
0.4\Gyr$ that is the characteristic time for Eddington-limited accretion mass
gain. In this section we use time normalized to Eddington time
$\tau=t/t_{Edd}$, where $\varkappa$ is Thomson (electron scattering) opacity. 
A general expression for Kerr parameter evolution is: 

\begin{equation}
\displaystyle\frac{da}{d\tau}=\displaystyle\frac{c}{G}\left(\displaystyle\frac{1}{M^2}\displaystyle\frac{dJ}{d\tau}-\displaystyle\frac{2aG}{Mc^3}\displaystyle\frac{dE}{d\tau} \right)
\end{equation}

\subsection{Accretion disk models}

In this subsection we give a small introduction to different accretion models and try to describe some of their limitations.

Standard disc model by \citet{SS73} now is the most widely used for intermediate accretion rates $\dot M \sim 0.01 - 1 \dot M_{Edd}$. In this model, there are at least three basic assumptions. One is ``alpha-prescription'', i.e. the assumption that viscous stress is proportional to thermal (gas+radiation) pressure with some coefficient $\alpha$ constant throughout the disk, $W_{r\phi}=\alpha p$. The second assumption is that accretion disk is geometrically thin that implies that the thermal velocity and the radial velocity are much smaller than Keplerian velocity, and all the energy dissipated by viscous processes is radiated locally. The thinness of the standard disc makes it difficult to simulate numerically. We use relativistic version of the model \citep{NT73} in our calculations. Third, less frequently mentioned, assumption is the boundary condition at the inner edge of the disk where the viscous stress is assumed to be zero. Near the last stable orbit, some of the standard disk assumptions are violated. Firstly, near the last stable orbit the radial velocity can become comparable to Keplerian. Another thing is viscous stress tensor may deviate from zero near the last stable orbit. In a more comprehensive analysis, boundary condition is non-trivial that may be viewed as a torque applied to the inner edge of the disc. 
It contributes to angular momentum exchange between accretion disk and black hole that may be effectively incorporated in the DML term.

For very large and very small accretion rates one should take into account advection process that makes accretion disk geometrically thick.  

For low accretion rates $\dot M\lesssim 0.01 \dot M_{Edd}$ we use ADAF model by \citet{NY95} that describes a geometrically thick disk with radial energy advection, that has some non-negligible  radial pressure gradient and radial velocity. Narayan\& Yi's model is essentially self-similar hence all the velocities are locally proportional to the Keplerian velocity. The self-similar nature is the main disadvantage of this ADAF model since boundary conditions cannot be satisfied. The other problem is its instability to thermal perturbations \citep{Blandford02}. This model can explain some observational results \citep{Mahadevan98}, but other observations and simulations indicate the insufficiency of ADAF approach \citep{Oda2012}. The main advantage of this model is its simplicity and the existence of analytical expressions for all the quantities such as gas pressure and the thickness of the disk. 

The other model applied for low accretion rates is a modification of ADAF, Advection-dominated inflow-outflow solution -- ADIOS \citep{Begelman99}. Apart from advection, this model also takes into account isotropic wind from the disk. Therefore, only a small fraction of the accreting matter falls onto the BH. Note however that for hydrodynamical disks, radiative inefficiency does not directly imply formation of massive outflows, as it was shown by \citet{Abramowicz00} using analytical arguments.

The only model that consistently considers the transonic nature of the flow near the last stable orbit is the slim disk by \citet{Abramowicz88}. Slim disk may be thought of as a standard disk where the assumptions of negligibility of pressure gradient and $({\mathbf v}\nabla){\mathbf v}$ terms in Euler equations are relaxed. This model can be applied to any accretion disk within a wide range of accretion rates. 
The main reason why we do not use this model is lack of analytical solutions for the equation set defining the slim disk.

 Supercritical accretion was considered with help of the model by \citet{lipunova99} and its modified version by \citet{poutanen07} . This model is a generalization of the Shakura-Synaev's model for super-Eddington  accretion rates.  Unlike the slim-disk approach, it takes into account the loss of accreting material in the disk wind but includes advection that is important for large accretion rates. Numerical simulations confirm existence of outflows (see \citet{Ohsuga05}, \citet{Fukue11}, \citet{Yang2014}). However, some recent numerical results \citep{Sadowski14} argue for a more complex structure of the flow where winds exist at distances larger than the classical non-relativistic spherization radius but mass accretion rate is practically constant with radius in the inner parts of the disk. This may imply that the real mass accretion rates are several times larger in the super-critical case. This allows black holes to grow more rapidly than expected in the na\"ive non-relativistic approach but does not alter significantly the expected spin-up curve that we find indistinguishable from the dust-like solution (see below section 5.2.3).

Surely, there is a lot of limitations and simplifications in the models we use. But we hope that does not contribute significantly to our results. In discussion we consider some possible consequences of the over-simplifications introduced by the models used in our work. 

\subsection{Standard disc}

Falling matter with high angular momentum forms an accretion disc around a black hole. As we have already mentioned above we used the relativistic version of Shakura-Sunyaev's model introduced by \citet{NT73}. This accretion regime apparently occurs in outbursts of X-ray novae and in the bright AGNs (Seyfert galaxies and QSOs).  In this model, the accretion disc is considered geometrically thin and optically thick and is divided into three zones A, B and C depending on the terms dominating pressure and opacity. Since we study the hottest inner parts of the accretion disc, only zones A and B were considered, where the main opacity  contribution is electron scattering. Gas and radiation pressure  dominate in zones B and A, respectively. 

\begin{equation}\label{pressure1}
p(r) = 
\left\{ \begin{array}{lc} 
1.93 \cdot 10^{16}  \alpha^{-1} m^{-1} r^{-3/2}  {\cal A}^{-2} {\cal B}^{2} {\cal S} ~ {\rm erg \, cm^{-3}} & \mbox{zone A}\\ 
2.65 \cdot 10^{18}  \alpha^{-9/10} m^{-17/10} {\dot m}^{4/5} r^{-51/20}  {\cal A}^{-1} {\cal B}^{1/5} {\cal D}^{-2/5} {\cal S}^{1/2} {\cal F}^{4/5} ~ {\rm erg \, cm^{-3}} & \mbox{zone B}\\
\end{array} \right.
\end{equation}
Here ${\cal A}, {\cal B}, {\cal C}, {\cal D}, {\cal F}, {\cal S}$ are relativistic correction factors (see \citet{penna}).\\

The boundary between these zones is determined by equality between radiation and gas pressures and lies at $r=r_{ab}$: 

\begin{equation}
r_{ab}=96 \alpha^{2/21} m^{2/21} {\dot m}^{16/21} {\cal A}^{20/21} {\cal B}^{-12/7} {\cal D}^{-8/21} {\cal S}^{-10/21} {\cal F}^{16/21}
\end{equation}

For a low accretion rates, effective radius is situated in zone B, and for
accretion rates higher than $\sim 0.2 \dot M_{Edd}$ (for $M=10^7 M\odot$) in
zone A.

In our calculations we used viscosity parameter $\alpha$ equal to 0.1 and dimensionless coefficient $\beta$ equal to 1. Variations of these parameters will be discussed in sec.~5.2.
 Accretion rate onto a black hole and its mass were normalized to the critical accretion rate (Eddington limit $L_{Edd}/c^2$) and to the solar mass, respectively: 

$$
\dot m=\displaystyle\frac{\dot M}{\dot M_{edd}}=\displaystyle\frac{\dot M c \kappa}{4\pi GM}
$$

$$
m=\displaystyle\frac{M}{M \odot}
$$

One can re-write the main evolutionary equations (14-17) for each of the two zones A and B in the case of standard relativistic sub-Eddington disc using eq. (18) for the pressure and Keplerian frequency as viewed from infinity $\Omega_{K}=\displaystyle\frac{c^3}{GM}(r_{eff}^{3/2}+a)^{-1}$ at effective radius $r_{eff}=1.3r_{in}$ as follows:

\subsubsection{Zone A}

\begin{equation}\label{E:vr45}
\left(\displaystyle\frac{dE}{d\tau}\right)_{em} = 
\left\{ \begin{array}{lc}
0.017  M_{\odot} c^2 \times \alpha^{-1} \beta^{-1}   a^2 r^2_{H} m {r_{eff}}^{-3/2} {\cal A}^{-2} {\cal B}^{2} {\cal S}  & \mbox{ BZ}\\
0.322  M_{\odot} c^2 \times \alpha^{-1} \beta^{-1} m {\dot m}^{2} {r_{eff}}^{1/2} (r_{eff}^{3/2}+a)^{-1} \displaystyle\left((r_{eff}^{3/2}+a)^{-1}-\frac{a}{2r_{H}}\right) {\cal A}^{2} {\cal B}^{-4} {\cal C}^{2} {\cal D}^{-2} {\cal S}^{-1} {\Phi}^{2} & \mbox{ DML}\\
\end{array} \right.
\end{equation}

\begin{equation}\label{E:vr4}
\left(\displaystyle\frac{dJ}{d\tau}\right)_{em} = 
\left\{ \begin{array}{lc}
0.07  \displaystyle\frac{GM_{\odot}^{2}}{c} \times \alpha^{-1} \beta^{-1}   a r^3_{H} m^{2}  {r_{eff}}^{-3/2} {\cal A}^{-2} {\cal B}^{2} {\cal S}  & \mbox{ BZ}\\ 
0.322  \displaystyle\frac{GM_{\odot}^{2}}{c} \times \alpha^{-1} \beta^{-1} m^{2} {\dot m}^{2} {r_{eff}}^{1/2} \displaystyle\left((r_{eff}^{3/2}+a)^{-1}-\frac{a}{2r_{H}}\right)  {\cal A}^{2} {\cal B}^{-4} {\cal C}^{2} {\cal D}^{-2} {\cal S}^{-1} {\Phi}^{2} & \mbox{ DML}\\
\end{array} \right.
\end{equation}

\begin{equation}\label{E:vr1}
\left(\displaystyle\frac{da}{d\tau}\right)_{em} = 
\left\{ \begin{array}{lc}
0.07  \alpha^{-1} \beta^{-1}   a r^3_{H} {r_{eff}}^{-3/2} \displaystyle\left(1-\frac{a^{2}}{2r_{H}}\right) {\cal A}^{-2} {\cal B}^{2} {\cal S}  & \mbox{ BZ}\\
0.322  \alpha^{-1} \beta^{-1}  {\dot m}^{2} {r_{eff}}^{1/2} \displaystyle\left((r_{eff}^{3/2}+a)^{-1}-\frac{a}{2r_{H}}\right) \displaystyle\left(1-2 a (r_{eff}^{3/2}+a)\right)  {\cal A}^{2} {\cal B}^{-4} {\cal C}^{2} {\cal D}^{-2} {\cal S}^{-1} {\Phi}^{2} & \mbox{ DML}\\
\end{array} \right.
\end{equation}

\subsubsection{Zone B}

\begin{equation}\label{E:vr}
\left(\displaystyle\frac{dE}{d\tau}\right)_{em} = 
\left\{ \begin{array}{lc}
2.1  M_{\odot} c^2 \times  \alpha^{-9/10} {\beta}^{-1} a^2 r^2_{H} m^{11/10} {\dot m}^{4/5} {r_{eff}}^{-51/20}  {\cal A}^{-1} {\cal B}^{1/5} {\cal D}^{-2/5} {\cal S}^{1/2} {\Phi}^{4/5}    & \mbox{ BZ}\\
0.0036 M_\odot c^2 \times \alpha^{-11/10} \beta^{-1} m^{9/10} {\dot m}^{6/5} r_{eff}^{31/20} (r_{eff}^{3/2}+a)\displaystyle\left((r_{eff}^{3/2}+a)^{-1}-\frac{a}{2r_{H}}\right) {\cal A} {\cal B}^{-11/5} {\cal C} {\cal D}^{-8/5} {\cal S}^{-1/2} {\Phi}^{6/5}  & \mbox{ DML}\\
\end{array} \right.
\end{equation}

\begin{equation}\label{E:vr}
\left(\displaystyle\frac{dJ}{d\tau}\right)_{em} = 
\left\{ \begin{array}{lc}
8.39  \displaystyle\frac{GM_{\odot}^{2}}{c} \times  \alpha^{-9/10} {\beta}^{-1} a r^3_{H} m^{21/10} {\dot m}^{4/5} {r_{eff}}^{-51/20}  {\cal A}^{-1} {\cal B}^{1/5} {\cal D}^{-2/5} {\cal S}^{1/2} {\Phi}^{4/5}    & \mbox{ BZ}\\
0.0036  \displaystyle\frac{GM_{\odot}^{2}}{c} \times \alpha^{-11/10} \beta^{-1} m^{19/10} {\dot m}^{6/5} {r_{eff}}^{31/20} \displaystyle\left((r_{eff}^{3/2}+a)^{-1}-\frac{a}{2r_{H}}\right)  {\cal A} {\cal B}^{-11/5} {\cal C} {\cal D}^{-8/5} {\cal S}^{-1/2} {\Phi}^{6/5} & \mbox{ DML}\\
\end{array} \right.
\end{equation}

\begin{equation}\label{E:vr}
\left(\displaystyle\frac{da}{d\tau}\right)_{em} = 
\left\{ \begin{array}{lc}
8.39 \alpha^{-9/10} {\beta}^{-1}  a r^3_{H} m^{1/10} {\dot m}^{4/5} {r_{eff}}^{-51/20} \left(1-\displaystyle\frac{a^2}{2r_{H}}\right)  {\cal A}^{-1} {\cal B}^{1/5} {\cal D}^{-2/5} {\cal S}^{1/2} {\Phi}^{4/5}     & \mbox{ BZ}\\
0.0036  \times \alpha^{-11/10} \beta^{-1} m^{-1/10} {\dot m}^{6/5} r_{eff}^{31/20} \displaystyle\left((r_{eff}^{3/2}+a)^{-1}-\frac{a}{2r_{H}}\right)  \displaystyle\left(1-2 a (r_{eff}^{3/2}+a)^{-1}\right) {\cal A} {\cal B}^{-11/5} {\cal C} {\cal D}^{-8/5} {\cal S}^{-1/2} {\Phi}^{6/5}  & \mbox{ DML}\\
\end{array} \right.
\end{equation}

Here BZ and DML denote Blandford-Znajek scenario and direct magnetic link, respectively.

\subsection{Advective disc}\label{sec:addisc}

For low accretion rates ($\mdot \lesssim 10^{-2}$), standard accretion disc is predicted to be unstable to evaporation \citep{meyer}. In this case, accretion flows are optically thin and geometrically thick and lose  inefficiently the energy released by viscous heating. In combination with efficient angular momentum transfer, this leads to a non-Keplerian flow where the heat is advected towards the black hole rather than lost with radiation. This type of accretion flows is known as Advection-Dominated Accretion Flows (ADAF).

Massive black holes in non-active (quiescent) galactic nuclei (like Sgr A*) in general show radiatively inefficient accretion having very little in common with the standard thin disc accretion picture.  

There are two main effects that characterize this accretion regime in the disc: radial velocities are non-Keplerian and the disc is geometrically thick.
Since there is no comprehensive relativistic analytical model for ADAF flows similar to the standard disc model, we use the non-relativistic self-similar model proposed by \citet{NY95}.
 In this model, all velocity components at some radius are proportional to the Keplerian velocity.

In particular, the radial velocity component is $v_r(R)= -\displaystyle\frac{(5+2\epsilon)}{3\alpha^2}\left(\left[1+\displaystyle\frac{18\alpha^2}{(5+2\epsilon)^2}\right]^{1/2}-1\right)v_{K}(R)= -c_1\alpha v_K(R)$, where $0<c_1<1$. Angular rotational frequency $\Omega(R)=\left(\displaystyle\frac{2\epsilon(5+2\epsilon)}{9\alpha^2}\left(\left[1+\displaystyle\frac{18\alpha^2}{(5+2\epsilon)^2}\right]^{1/2}-1\right)\right)^{1/2}\displaystyle\frac{v_{K}}{R}=c_2\displaystyle\frac{v_K}{R}$.
 Speed of sound squared $c^2_s(R)=\displaystyle\frac{2(5+2\epsilon)}{9\alpha^2}\left(\left[1+\displaystyle\frac{18\alpha^2}{(5+2\epsilon)^2}\right]^{1/2}-1\right)v_{K}(R)= c_3 v^2_K(R)$. 

In our simulations we  use $\alpha=0.1$ and $\epsilon=1$, implying $c_1=0.43$, $c_2=0.53$ and $c_3=0.285$. 

One more correction should be made connected to the high temperature of the accreted gas.
One can estimate dimensionless specific enthalpy as follows: 

\begin{equation}
\mu=1+\displaystyle\frac{U+\Pi}{\Sigma c^2}=1+\frac{5}{4}\frac{\Pi}{\Sigma c^2}=1+\frac{5}{4}\frac{p}{\rho c^2}=1+\displaystyle\frac{5}{4}\displaystyle\frac{c_3}{r}
\end{equation}

Here, $\Sigma$ and $\Pi$ are vertically integrated mass density $\rho$ and
pressure $p$.
Black hole spin evolution in absence of electromagnetic terms for ADAF regime is described by the following equation:

\begin{equation}
\displaystyle\frac{da}{dt}=\mu \displaystyle\frac{\dot M}{M}\left(c_2 j^{\dagger}_K(r_{in})-2a E^{\dagger}(r_{in}) \right)
\end{equation}

The main evolutionary equations for the case of conventional advective flow are the following: 
\begin{equation}\label{E:vmr}
\left(\displaystyle\frac{dE}{d\tau}\right)_{em} = 
\left\{ \begin{array}{lc}
\displaystyle\frac{\sqrt{5}}{320}  M_{\odot} c^2 \times \alpha^{-1} {\beta}^{-1} c^{-1}_{1} c^{1/2}_{3} a^2 r^2_{H} r^{-5/2}_{eff} m \dot m  & \mbox{ BZ}\\
\displaystyle\frac{\sqrt{5}}{8}  M_{\odot} c^2 \times  \alpha^{-1} {\beta}^{-1}  c^{-1}_{1} c^{3/2}_{3}\Omega_{*} m r^{3/2}_{eff}
\displaystyle\left(\Omega_{*}-\frac{a}{2r_{H}}\right)\dot m & \mbox{ DML}\\
\end{array} \right.
\end{equation}

\begin{equation}\label{E:vr3}
\left(\displaystyle\frac{dJ}{d\tau}\right)_{em} = 
\left\{ \begin{array}{lc}
\displaystyle\frac{\sqrt{5}}{80}  \displaystyle\frac{GM_{\odot}^{2}}{c}  \times \alpha^{-1} {\beta}^{-1} c^{-1}_{1} c^{1/2}_{3} a  r^3_{H} r^{-5/2}_{eff} m^2 \dot m  & \mbox{ BZ}\\
\displaystyle\frac{\sqrt{5}}{8}  \displaystyle\frac{GM_{\odot}^{2}}{c}  \times \alpha^{-1} {\beta}^{-1} c^{-1}_{1} c^{3/2}_{3}\displaystyle\left(\Omega_{*}-\frac{a}{2r_{H}}\right){r_{eff}}^{3/2} m^2 \dot m & \mbox{ DML}\\
\end{array} \right.
\end{equation}

\begin{equation}\label{E:vrw}
\left(\displaystyle\frac{da}{d\tau}\right)_{em} = 
\left\{ \begin{array}{lc}
\displaystyle\frac{\sqrt{5}}{80}\alpha^{-1} \beta^{-1} c^{-1}_{1} c^{1/2}_{3} a r^3_{H} r^{-5/2}_{eff} \dot m \left(1-\displaystyle\frac{a^{2}}{2r_{H}}\right)  & \mbox{ BZ}\\
\displaystyle\frac{\sqrt{5}}{8} \alpha^{-1} \beta^{-1} c^{-1}_{1} c^{3/2}_{3}\displaystyle\left(\Omega_{\ast}-\frac{a}{2r_{H}}\right) {r_{eff}}^{3/2} \left(1-2a \Omega_{\ast}\right) \dot m & \mbox{ DML}\\
\end{array} \right.
\end{equation}

Here $\Omega_{*}=c_2\Omega_K$ is angular frequency for ADAF disc. The pressure is calculated in the following way:

\begin{equation}\label{pressure2}
p=1.7\times 10^{16} \alpha^{-1} c_1^{-1} c_3^{1/2} \dot m m^{-1} r^{-5/2}_{eff} {\rm erg\, cm^{-3}}
\end{equation}

 There are some unconstrained coefficients and it is important to
  understand how much Kerr parameter evolution depends on their
  values. Numerical coefficient $\epsilon$ depends on the ratio of specific
  heats $\gamma$ and on the fraction of advected energy $f$ as
  $\epsilon=\displaystyle\frac{1}{f}\displaystyle\frac{5/3-\gamma}{\gamma-1}$. In
  non-relativistic case $\gamma=5/3$, while for relativistic flow $\gamma \to
  4/3$. Fully advective disc has $f=1$. If $f\ll 1$, a large fraction of
  energy is locally radiated and the disc rapidly becomes thin. Hence we
  always use $f=1$ for ADAF.

Numerical coefficients $c_1$ and $c_3$ depend weakly on parameters $\epsilon$
and $\alpha$. In the range $0.1<\epsilon<1$, and $0.01<\alpha<1$, they vary by
about 10\%. Non-Keplerianity  parameter $c_2$ depends on $\epsilon$ and
$\alpha$ much stronger. Near the last stable orbit, no inward-directed
pressure gradient can be present hence the self-similar model probably
under-estimates the value of $c_2$. 
 Matter-only term is proportional to $c_2$, electromagnetic term is
 proportional to $\displaystyle\frac{c_2}{c_1}c_3^{3/2}$ that varies by about
 18\% in DML case and to $\displaystyle\frac{c_3^{1/2}}{c_1}$ that varies by
 about 21\% in BZ case, hence its variations due to unknown equation of state
 and viscosity do not exceed $\sim 0.2$. If we fix $c_2$ equilibrium Kerr
 parameter varies up to 0.1\% for BZ case and up to 6.5\% in DML case. The
 largest uncertainty is connected to the unknown value of $c_2$. Hopefully,
 further, more comprehensive studies will help constrain its value.

\subsection{Supercritical disc $\dot m>>1$}

When the accretion rate is not very low and not very high, the standard disc
theory can be applied. If accretion disc luminosity reaches the Eddington
limit, $H/R$ becomes close to 1 in the inner parts of the disc and the thin
disc approximation breaks down. There are at least two effects influencing the
properties of supercritical accretion: outflow formation (considered in
\citep{SS73}) and photon trapping (first considered in the advective Polish
doughnut model \citep{abram05}).

Effect of advection dominates when accretion rate is very high $\dot
m>10^3$. { Since there is numerical support for the outflow scenario (see for
example \citet{osuga07}), we use an
outflow-based model by \citet{lipunova99}.} An outflow starts at the
spherization radius where $H/R=1$. Below we will adopt spherization radius in
the form $r_{sp}=5/3 \dot m$.  If effective radius is less than spherization
radius, accretion rate becomes smaller and effective $\dot m_{eff}$ may be
written as follows (see \citet{lipunova99} and \citet{poutanen07}): 

\begin{equation}
\dot m_{eff}=
\left\{\begin{array}{lc}
\dot m\displaystyle\frac{r}{r_{sp}}\displaystyle\frac{1+2/3r^{-5/2}}{1+2/3r_{sp}^{-5/2}}  & \mbox{$r<r_{  sp}$}\\
\dot m  & \mbox{$r>r_{sp}$}\\
\end{array}\right.
\end{equation}

The matter flowing from the region inside the spherization radius removes angular momentum. Radial flux of angular momentum is:

\begin{equation}
g(r)=
\left\{\begin{array}{lc}
\displaystyle\frac{\dot m r^{3/2}}{3r_{sp}}\displaystyle\frac{1-r^{-5/2}}{1+2/3r_{sp}^{-5/2}}  & \mbox{ $r<r_{sp}$}\\
\displaystyle\frac{\dot m \sqrt{r}}{3}\displaystyle\frac{1-r_{sp}^{-5/2}}{1+2/3r_{sp}^{-5/2}}+\dot m(\sqrt{r}-\sqrt{r_{sp}})  & \mbox{$r>r_{sp}$}\\
\end{array}\right.
\end{equation}

Taking into account the features of supercritical regime, mass and spin evolution is described by the following equations:
 
\begin{equation}\label{E:vr45}
\left(\displaystyle\frac{dE}{d\tau}\right)_{em} = 
\left\{ 
\begin{array}{lc}
0.02  M_{\odot} c^2 \times \alpha^{-1} \beta^{-1}   a^2 r^2_{H} {r_{eff}}^{-1} {\dot m_{eff}} r_{in}^{-1/2} m {g(r_{eff})}^{-1} \displaystyle\frac{1-\left(\displaystyle\frac{r_{eff}}{r_{in}}\right)^{-5/2}}{1+\displaystyle\frac{2}{3}\left(\displaystyle\frac{r_{eff}}{r_{in}}\right)^{-5/2}}  & \mbox{ BZ}\\
0.21  M_{\odot} c^2 \alpha^{-1} \beta^{-1} {\dot m_{eff}} ({r_{eff}}^{3/2}+a)^{-1} \displaystyle\left(({r_{eff}}^{3/2}+a)^{-1}-\frac{a}{2r_{H}}\right) g(r_{eff}) m r_{in}^{1/2}  \displaystyle\frac{1-\left(\displaystyle\frac{r_{eff}}{r_{in}}\right)^{-5/2}}{1+\displaystyle\frac{2}{3}\left(\displaystyle\frac{r_{eff}}{r_{in}}\right)^{-5/2}} & \mbox{ DML}\\
\end{array} 
\right.
\end{equation}

\begin{equation}\label{E:vr4}
\left(\displaystyle\frac{dJ}{d\tau}\right)_{em} = 
\left\{ 
\begin{array}{lc} 
0.08 \displaystyle\frac{GM_{\odot}^{2}}{c}  \times \alpha^{-1} \beta^{-1}   a r^3_{H} {r_{eff}}^{-1} {\dot m_{eff}} m^2 r_{in}^{-1/2} {g(r_{eff})}^{-1} \displaystyle\frac{1-\left(\displaystyle\frac{r_{eff}}{r_{in}}\right)^{-5/2}}{1+\displaystyle\frac{2}{3}\left(\displaystyle\frac{r_{eff}}{r_{in}}\right)^{-5/2}}  & \mbox{ BZ}\\
0.21  M_{\odot} c^2 \alpha^{-1} \beta^{-1} {\dot m_{eff}} \displaystyle\left(({r_{eff}}^{3/2}+a)^{-1}-\frac{a}{2r_{H}}\right) g(r_{eff}) m^2 r_{in}^{1/2}  \displaystyle\frac{1-\left(\displaystyle\frac{r_{eff}}{r_{in}}\right)^{-5/2}}{1+\displaystyle\frac{2}{3}\left(\displaystyle\frac{r_{eff}}{r_{in}}\right)^{-5/2}} & \mbox{ DML}\\
\end{array} 
\right.
\end{equation}

\begin{equation}\label{E:vr1}
\left(\displaystyle\frac{da}{d\tau}\right)_{em} = 
\left\{\begin{array}{lc}
0.08 \alpha^{-1} \beta^{-1}   a  r^3_{H} {r_{eff}}^{-1}  {\dot m_{eff}} r_{in}^{-1/2}  \displaystyle\left(1-\frac{a^2}{2 r_{H}}\right) {g(r_{eff})}^{-1} \displaystyle\frac{1-\left(\displaystyle\frac{r_{eff}}{r_{in}}\right)^{-5/2}}{1+\displaystyle\frac{2}{3}\left(\displaystyle\frac{r_{eff}}{r_{in}}\right)^{-5/2}}   & \mbox{ BZ}\\

0.21  \alpha^{-1} \beta^{-1} {\dot m_{eff}} r_{in}^{1/2} \displaystyle\left((r_{eff}^{3/2}+a)^{-1}-\frac{a}{2r_{H}}\right) \displaystyle\left(1-2 a (r_{eff}^{3/2}+a)^{-1}\right) g(r_{eff}) \displaystyle\frac{1-\left(\displaystyle\frac{r_{eff}}{r_{in}}\right)^{-5/2}}{1+\displaystyle\frac{2}{3}\left(\displaystyle\frac{r_{eff}}{r_{in}}\right)^{-5/2}}    & \mbox{ DML}\\
\end{array} \right.
\end{equation}

\section{Magnetic field decay}\label{sec:decay}

\subsection{Hall decay}

In an accretion disc magnetic field decay is in equilibrium with  magnetic field amplification by the instabilities such as MRI \citep{MRI} and dynamo processes \citep{dynamo}. This equilibrium state is shifted inside the last stable orbit where the plasma becomes  magnetically dominated (gas and radiative pressure are reduced but magnetic field stress remaines practically unchanged). In these conditions magnetic field decay is primarily due to Hall cascade \citep{gold}. This effect does not change the total magnetic energy, but it can  transfer the energy to smaller spatial scales where Ohmic decay runs faster. 
We assume that magnetic field lines, connecting a black hole with the disc have a curvature radius $L$ of the order $GM/c^2$.
Following \citet{gold} we can write magnetic field evolution equation for magnetic field affected by Hall effect: 

\begin{equation}\label{max1}
\displaystyle\frac{\partial \vector B}{\partial t}=\displaystyle\frac{-c}{4\pi n e}\nabla \times [(\nabla \times \vector B)\times \vector B]+\displaystyle\frac{c^2}{4\pi \sigma}\nabla^2 \vector B
\end{equation}

The first term in this equation describes Hall cascade.

For the largest loops, one may estimate $\partial \vector B/\partial t\sim B/\tau$,
$\nabla \sim 1/L$ and obtain Hall timescale as following:

\begin{equation}\label{max3}
\tau_{Hall}=\displaystyle\frac{4\pi n e L^2}{cB}
\end{equation}

Here $n$ is electron concentration and $L$ is characteristic curvature scale. For geometrically thick flow electron concentration may be estimated as $n\simeq \displaystyle\frac{\dot m}{\sigma_T}\left(\displaystyle\frac{c^2}{GM}\right)$.

Characteristic Hall timescale is:

\begin{equation}
\tau_{Hall}=\displaystyle\frac{4\pi \dot m c e R_H^2}{\sigma_T GM B} =
\dfrac{4\pi e}{\sigma_T B} \mdot r_H^2 \times \tau_{dyn}
\end{equation}

Hall decay becomes important when $\tau_{Hall} \lesssim t_{dyn}$, when mass
accretion rate is very small. Using
expressions for an advective disc (section \ref{sec:addisc}), one arrives to
the following estimate:

$$
\tau_{Hall}\sim 10^7 \sqrt{\mdot m} \times \tau_{dyn}
$$

Hall drift is thus unimportant for astrophysical black holes. 

The second term in  eq. (\ref{max1}) describes Ohmic decay and its timescale $\tau_{Ohm}\propto \sigma$, but conductivity of ionized relativistic plasma is very large, hence Ohmic timescale approaches infinity. 

\subsection{Joule losses}\label{subdecay}

In Blandford-Znajek case magnetic field lines connect the black hole horizon with a distant load. Magnetic flux is conserved, hence large-scale magnetic field having non-zero flux through the equatorial plain can not dissipate and accumulates up to equipartition. These magnetic fields may be supported by electric currents in the accretion disc where resistance and dissipation are negligibly low.{ When we consider classical Blandford-Znajek magnetic field configurations, there is always a non-zero magnetic flux through the equatorial plane.
 As long as magnetic field lines are not allowed to move outwards (that is true as long as the field is frozen-in and accretion disc is present), the magnetic flux through the region inside
 the last stable orbit is also conserved, and one can set a lower limit for the magnetic field strength. Hence in the BZ case, magnetic flux conservation prohibits magnetic field dissipation below  certain level.}
In the direct magnetic link case there are always tangential magnetic fields near the black hole horizon and the currents should at least partially flow close to the horizon where effective resistance appears due to general relativity effects. Hence we will consider magnetic field decay effect only for the DML case.
 
Classical DML scenario involving magnetic loops extending above the disc is
excluded by numerical simulations (such as \citet{Shafee10}). 
Instead, most of the magnetic field energy is stored in frozen-in
loops that are advected with the falling matter. 
If magnetic fields are frozen into the free-falling matter inside the ISCO,
the direction of the field should alter at least once at any radius, even
close to the horizon.  Magnetic field configuration is that of a driven
current sheet, where dissipation is concentrated towards the middle plane and
may be interpreted as magnetic field reconnection.   Magnetic field dissipation
leads to a magnetic torque value efficiently altered by some factor of $\chi <
1$. 

To estimate the magnetic field decay rate in membrane paradigm, one may
consider toroidal currents responsible for radial magnetic field direction
change. The amplitude of such a current is, approximately:

$$
j_\varphi = \frac{c}{4\pi} [\nabla B] \simeq \frac{c}{4\pi} \frac{B^r}{D},
$$

where $D$ is the vertical scale of the falling magnetized flow. The radial
extent of the flow is about $R_{in}-R_{H}$. Effective surface current is
thus:

$$
g_\varphi \simeq  \frac{c}{4\pi} \frac{R_{in}-R_H}{D} B^r
$$

Magnetic field $B^r$ here is estimated at the stretched horizon and differs
from the ``initial'' magnetic field near the ISCO (that is close to
equipartition) by the factor of disc cross-section ratio $\dfrac{H_{in} R_{in}}{DR}$
Magnetic field energy dissipated near the horizon may be estimated as the
following integral upon the BH horizon:

$$
-\dot{E}_H \simeq \frac{4\pi}{c} \int g_\varphi^2 dS
$$

$$
-\dot{E}_H \simeq 4\pi RH \times \left(\frac{c}{4\pi} \frac{R_{in}-R_H}{D}
B^r\right)^2 \times \frac{4\pi}{c} = -8\pi \left(\dfrac{H_{in}}{D}\right)^2
\dfrac{r_{in}^2(r_{in}-r_H)^2}{r_H^2} \times \left(\frac{GM}{c^2}\right)^2 p_M c
$$

This is the estimated loss of the magnetic field energy near the horizon. Note
that it
does not involve any processes connected to black hole rotation.
Magnetic field energy input from the disc: 

$$
\dot{E}_D \simeq p_M \times u^r \times 4\pi H_{in} R_{in} \simeq 
p_M \times c \alpha \dfrac{H_{in}}{R_{in}} \times 4\pi H_{in} R_{in} = 
4\pi \alpha \left(\dfrac{H_{in}}{R_{in}}\right)^3 \dfrac{r_{in}}{r_{in}^{3/2}+a} \times
\left(\frac{GM}{c^2}\right)^2 p_M c
$$

Practically always $-\dot{E}_H \gg \dot{E}_D$, and the two rates become
comparable only if viscosity is high and the flow is geometrically
thick. Equilibrium between field decay and replenishment requires $H_{in}/R_{in}\gtrsim 1$. 

Presence of a decay mechanism
may mean that the magnetic field responsible for radial transfer is generally
smaller than one may expect. Inner flow thickness $D$ close to the horizon may
be estimated as $D\simeq H_{in}$

if the flow geometry is not altered strongly by magnetic field decay itself.

$$
\chi \simeq \exp\left( -t_{replenishment} / t_{decay} \right) \simeq
\exp\left( \dot{E}_H / \dot{E}_D \right)
$$

Substituting all the quantities:

$$
\chi \simeq \exp\left(
-\dfrac{2 (r_{in}-r_H)^2 r_H}{\alpha h_{in} d^3}\right)
$$
where $H_{in}=h_{in}GM/c^2$ and $D=d GM/c^2$.

\begin{figure}
\centering
\includegraphics[width=1.0\columnwidth]{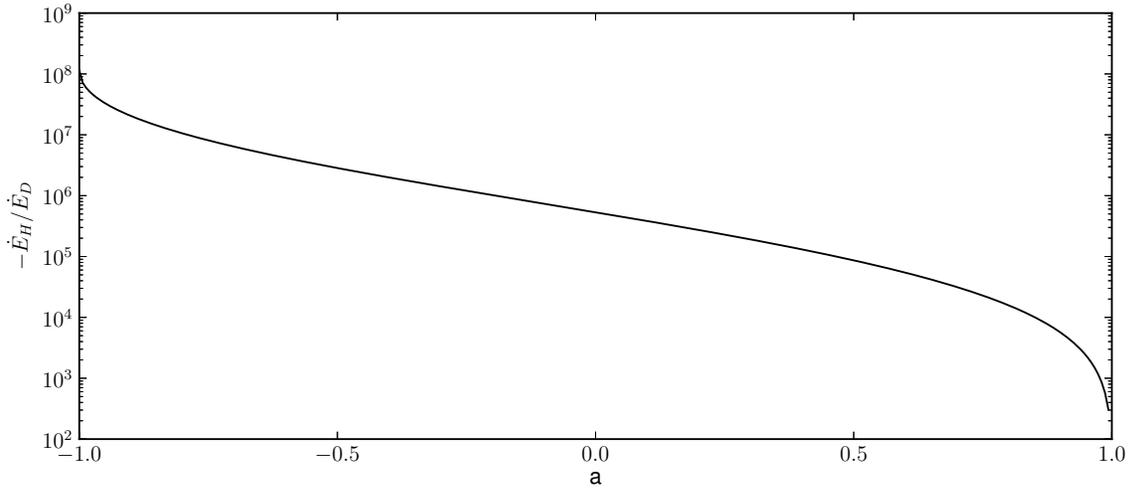}
\caption{Ratio of magnetic field decay $\dot{E}_H$ and replenishment
  $\dot{E}_D$ rates as a function of Kerr parameter. Viscosity parameter
  $\alpha=0.1$, disc initial thickness $H_{in}=0.3R_{in}$. }\label{fig:edeh}
\end{figure}

Figure \ref{fig:edeh} shows the estimated decay factor as a function of Kerr
parameter.

Summarizing, magnetic field decay is an important effect whenever there is
tangential magnetic field component anywhere on the stretched horizon. Since
magnetic field decay rate is generally several orders of magnitude larger than
the rate of magnetic field replenishment from the disc (primarily because of
$v_r \ll 1$), we conclude that DML momentum transfer is strongly damped.

\section{Results} \label{sec:results}

\subsection{Equilibrium Kerr parameter}

Spin evolution equation may be written as:

\begin{equation}\label{ono}
\dfrac{da}{d\tau} = \mu \mdot \left[ j^\dagger - 2a E^\dagger \right]+
\dfrac{c}{G} \left[ \dfrac{1}{M^2} \left(\dfrac{dJ}{d\tau}\right)_{em} -
  \dfrac{2aG}{Mc} \left(\dfrac{dM}{d\tau}\right)_{em}\right] \times \chi
\end{equation}

First term here describes the Kerr parameter evolution connected to the
angular momentum advected by the falling matter. It is proportional to mass
accretion rate and usually positive. The second term corresponds to the
contribution of electromagnetic processes described in sec. 2.  This term is
always negative in Blandford-Znajek regime and changes sign in the DML case
when corotation radius equals the effective disc radius. 

Additional multiplier $\chi$ takes into account magnetic field decay (see section~\ref{sec:decay}).

As long as the first term is higher than the second by its absolute value, black hole is spun up toward $a\sim 1$ for a standard disc and up towards $a\sim 0.6$ for a sub-Keplerian ADAF disc. Here we used $j^\dagger=c_2j_{K}(r_{in})$, where $c_2$ is a parameter taking into account deviation from Keplerian law (see section~\ref{sec:addisc}) and $j_{K}(r_{in})$ is net Keplerian angular momentum at the inner radius.
 If the second term is balanced by the first, black hole spin-up stops at the equilibrium Kerr parameter value $a_{eq}<0.998$. 

\begin{figure}
\centering
\includegraphics[width=0.7\columnwidth]{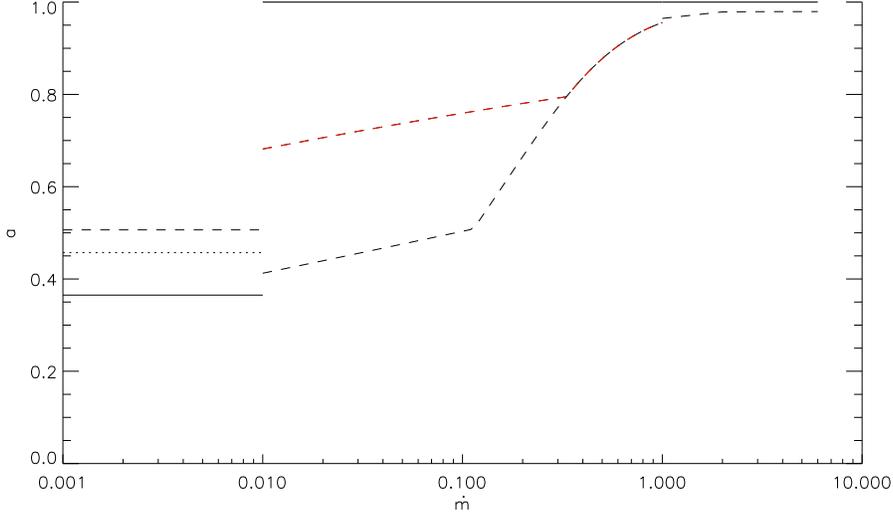}
\caption{Equilibrium Kerr parameters for different accretion rates are plotted
  for BZ processes by dashed lines and for DML processes without and with
  magnetic field decay by solid and dotted lines respectively. Black lines correspond to SMBH ($M_0=10^7 M\odot$) while red (grey)
  lines to the stellar mass black hole ($M_0=10 M\odot$) case.}\label{fig:aeq}
\end{figure}

{ 
Equilibrium spin parameter dependence on accretion rate is shown in Figure~\ref{fig:aeq}. Solid lines correspond to the DML case and BZ is plotted by a dashed line. 
Red (grey) lines are hereafter used for stellar-mass black holes ($M=10\Msun$) and black lines for SMBH ($M=10^7\Msun$). In this figure equilibrium Kerr parameter for DML processes with magnetic field decay is plotted by dotted line (see sections 4 and 5.2.1). 
In ADAF regime both electromagnetic and accretion terms are proportional to accretion rate $\dot m$ hence the equilibrium Kerr parameter does not depend on $\dot m$. 
Also it does not depend on the BH mass that is why the solutions for
supermassive and stellar-mass BHs are identical.

 For the standard disc case, the breakpoint marks transition from B to A zones. For the effective radius in zone B, spin evolution  strongly depends on BH mass, and the influence of the BZ process for SMBH is more significant than for LMBH. The boundary between A and B zones depends on mass and accretion rate. For LMBH this transition takes place for accretion rates greater than for SMBH. In zone A, the dependence on mass is insignificant and spin evolution proceeds in the similar way for the two types of BH.  But electromagnetic term depends on $\dot m$ somewhat weaker than accretion term, therefore the influence of BZ processes decreases with increasing $\dot m$. DML in inefficient in the Standard disc case for both SMBH and LMBH.
}

Equation (41) has no more than one attractor in the range $-1<a<1$. For some parameter sets (DML+standard disc), there is no stable solution of the $da/dt=0$ equation and the BH is spun up until $a\simeq 0.998$ where other effects not considered in this work (such as radiation capture, see \citet{thorne74}) come into play. In other cases, the black hole is always spun up for $a<a_{eq}$ and spun-down for $a>a_{eq}$.

\subsection{Rotational evolution of a typical SMBH}

For a more detailed analysis of black hole rotational evolution we made simulations for different accretion regimes for a supermassive Kerr black hole with the following parameters: initial mass $M_0=10^7 M_{\odot}$, initial Kerr parameters $a_0=0$ and $a_0=0.7$. For arbitrary dimensionless
accretion rate $\dot{m}$, both mass and spin change at
characteristic timescales $\sim t_{Edd} / \dot{m}$.

For a constant
non-zero accretion rate, mass grows monotonically and may be used as an
independent variable. 

We consider ADAF regime with accretion rate $\dot m=10^{-3}$, standard disc with $\dot m=0.4$ and supercritical regime with $\dot m=10$. We set magnetization parameter $\beta=1$ and $\alpha=0.1$. Note that if viscosity is created by chaotic magnetic field stresses $\alpha \sim \displaystyle\frac{B_{\phi}B_{r}}{4\pi p} \sim \displaystyle\frac{2}{\beta}$ hence $\alpha \beta=$const.  Since the two parameters enter the equations in combination $\alpha\beta$ (save the standard disc zone B where this statement is still approximately fulfilled; see section 3), the result depends weakly on the exact value of $\beta$.

\subsubsection{Rotational evolution in ADAF regime}

\begin{figure}
\centering
\includegraphics[width=0.7\columnwidth]{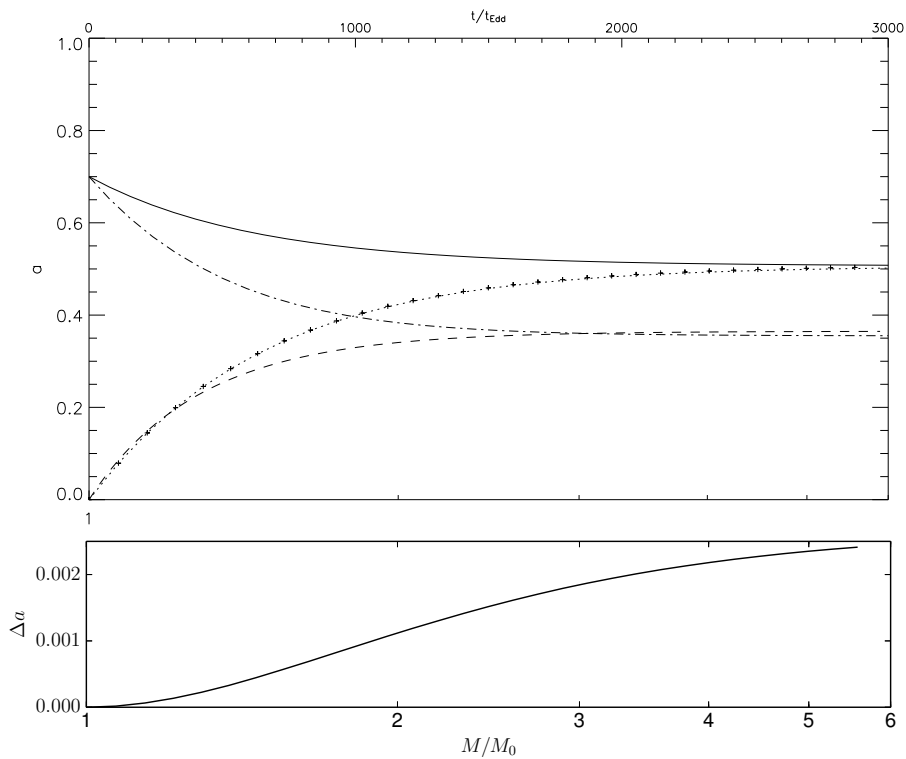}
\caption{Spin evolution of a black hole with initial mass $M_0=10^7
  M_{\odot}$, initial Kerr parameters $a_0=0$ and $a_0=0.7$, accretion rate
  $\dot M=10^{-3}\dot M_{Edd}$ (ADAF regime). Unmagnetized case evolution is
  plotted by crosses, evolution affected by DML processes is plotted by dashed
  ($a_0=0$) and dot-dashed ($a_0=0.7$) lines. Evolution affected by BZ process
  by a dotted line for $a_0=0$ and by a solid line for $a_0=0.7$. The bottom panel shows difference in Kerr parameter between the unmagnetized (Bardeen) and Blandford-Znajek cases for $a_0=0$. }\label{fig:adaf}
\end{figure}

In section \ref{sec:rev} it was noted that spin evolution under the matter accretion contribution for ADAF regime differs from the Bardeen solution due to non-Keplerian velocity law. 
In Fig.~\ref{fig:adaf}, spin evolution for the matter-only solution is plotted with
crosses, and one can see that there is an equilibrium Kerr parameter value
less than unity. Spin evolution affected by the Blandford-Znajek process is
plotted by a dotted line for $a_0=0$ and by a solid line for $a_0=0.7$. The
dashed and dot-dashed lines show the DML case for $a_0=0$ and $a_0=0.7$,
respectively . Maximal difference between BZ and sub-Keplerian matter-only
solution is about $0.0067$. We adopted $\epsilon^\prime=0.5$ that implies
parameter values of $c_1\simeq 0.43$, $c_2\simeq 0.53$ and $c_3\simeq 0.285$. 

Electromagnetic term for DML is higher than for BZ due to the difference in
numerical coefficients and in the different dependence on the effective
radius. Electromagnetic terms in these two cases differ by three orders of
magnitude, the DML process being more efficient. Black hole mass growth is
very slow and the Kerr parameter can not increase significantly during the
cosmological time if $\mdot \lesssim 10^{-2}$.

Note that the character of black hole rotation evolution at the timescales
longer than the evolutionary time scale $t_{Edd}/\dot m$ is insensitive to the
initial Kerr parameter. 

As it was shown above (in sec.4) magnetic field decay is efficient only in the
DML case. This process affects only the behaviour of the electromagnetic term
in the evolutionary equation. Influence of magnetic field decay is shown in
Fig. 5. Here evolution for matter contribution is plotted by a crosses,
evolution affected by DML processes without and with magnetic field decay is
plotted by a dotted and a solid lines, respectively. One can see that
equilibrium Kerr parameter increases from $0.36$ to $0.45$ due to the magnetic field
decay.

\begin{figure}
\centering
\includegraphics[width=0.7\columnwidth]{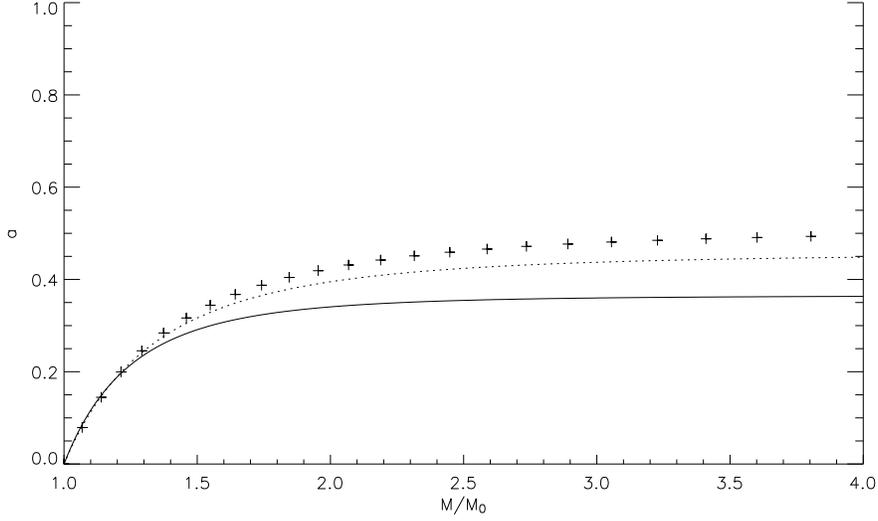}
\caption{Spin evolution of a black hole with initial mass $M_0=10^7
  M_{\odot}$, initial Kerr parameter $a_0=0$ accretion rate $\dot
  M=10^{-3}\dot M_{Edd}$. Evolution for accretion matter contribution is
  plotted by crosses, evolution affected by DML processes with and without
  magnetic field decay are plotted by dotted line and a solid line respectively.}\label{fig:stand}
\end{figure}

\begin{figure}
\centering
\includegraphics[width=0.7\columnwidth]{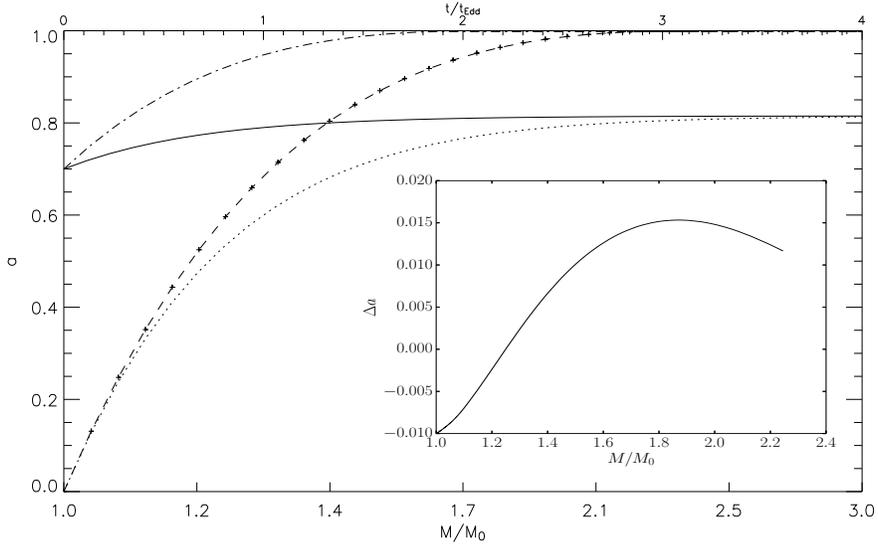}
\caption{Spin evolution of a black hole with initial mass $M_0=10^7 M_{\odot}$, initial Kerr parameters $a_0=0$ and $a_0=0.7$,  accretion rate $\dot M=0.4\dot M_{Edd}$ (Standard disc regime). Evolution for accretion matter contribution is plotted by crosses, evolution affected by DML processes is plotted by dashed ($a_0=0$) and dot-dashed ($a_0=0.7$) lines. Evolution affected by BZ process  by a dotted line for $a_0=0$ and by a solid line for $a_0=0.7$. The inset shows difference in Kerr parameter between the unmagnetized (Bardeen) and DML cases for $a_0=0$. }\label{fig:stand}
\end{figure}

\begin{figure}
\centering
\includegraphics[width=0.9\columnwidth]{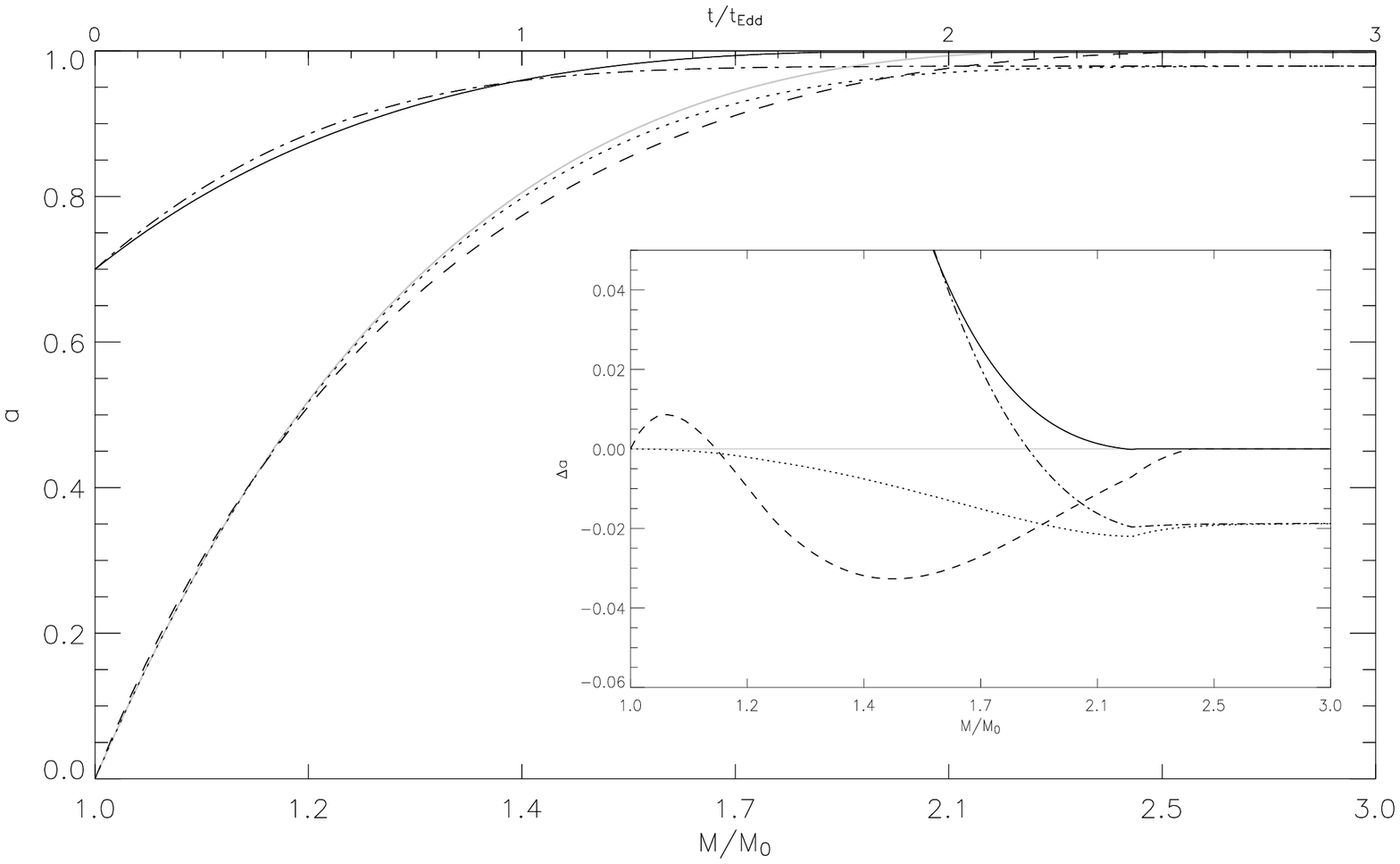}
\caption{Spin evolution of a black hole with initial mass $M_0=10^7
  M_{\odot}$, initial Kerr parameters $a_0=0$ and $a_0=0.7$, accretion rate
  $\dot M=3\dot M_{Edd}$ (Supercritical regime). Evolution for accretion
  matter contribution is plotted by grey solid line, evolution affected by DML
  processes is plotted by dotted ($a_0=0$) and solid ($a_0=0.7$)
  lines. Evolution affected by BZ process is shown by a dashed line for
  $a_0=0$ and by a dot-dashed line for $a_0=0.7$.}\label{fig:scrit}
\end{figure}

\subsubsection{Rotational evolution in standard disc regime}

The results for the standard accretion disc are presented in Fig.~\ref{fig:stand}. Here again spin evolution in the Bardeen's case is shown by crosses, spin evolution affected by the Blandford-Znajek processes is plotted by dotted ($a_0=0$) and solid ($a_0=0.7$) lines and in DML case by dashed ($a_0$) and by dot-dashed ($a_0=0.7$) lines. In this case direct magnetic link is much less efficient than Blandford-Znajek processes. 

Depending on magnetic field geometry, spin evolution may result either in fast
rotation with $a\simeq 1$ or in an intermediate $a\simeq a_{eq}$ where matter
spin-up is balanced by magnetic field spin-down. As in ADAF case, equilibrium Kerr parameter does not depend on the initial Kerr parameter value.

\subsubsection{Rotational evolution in supercritical regime}

In Fig.~\ref{fig:scrit} we present our results for the case of supercritical
regime. Here spin evolution of a black hole for matter contribution is plotted
by a grey solid line, evolution affected by DML processes is plotted by dotted ($a_0=0$) and solid ($a_0=0.7$) lines. Evolution affected by BZ process  by a dashed line for $a_0=0$ and by a dot-dashed line for $a_0=0.7$. The corrections to all Kerr parameters introduced by different electromagnetic processes are plotted in the inset by the same linestyles.         

\bigskip

As it was shown above (in sec.~\ref{sec:decay}) magnetic field decay is
efficient only in the DML case. This process affects only the behaviour of the
electromagnetic term in the evolutionary equation ~(41). 

Note that the character of black hole rotational evolution at the timescales longer than the evolutionary time scale $t_{Edd}/\mdot$ is insensitive to initial Kerr parameter.

\subsection{Black holes in X-ray binaries. Comparison with observational data}\label{sec:stellar}

X-ray binaries are usually divided into low-mass X-ray binaries (LMXB) with a donor star of several solar masses or smaller and high-mass X-ray binaries (HMXB) with the donor masses of tens of solar masses. 
In this section we consider spin evolution of such systems including X-ray novae and microquasars. Predictions of simple constant mass accretion rate models are compared to the observational data on several black hole X-ray binaries. 
 We assume that initial Kerr parameters of these objects are $a_0=0$ since
 there are reasons to expect BHs to be born slowly rotating \citep{moreno11}. According to the  work by \citet{Podsiadlowski}, accretion rates in these systems $\dot m \gtrsim 1$ hence we used a constant accretion rate model with $\dot m=1$ in our calculations.  We plot evolutionary tracks for black holes with masses 5, 7 and 10 $M_{\odot}$ in Figure 8. Solid and dashed lines represent the black hole evolution affected by Blandford-Znajek and DML processes, respectively. Here, DML is indistinguishable from Bardeen solution.

\begin{figure}\label{fig:aevolv}
\centering
\includegraphics[width=0.7\columnwidth]{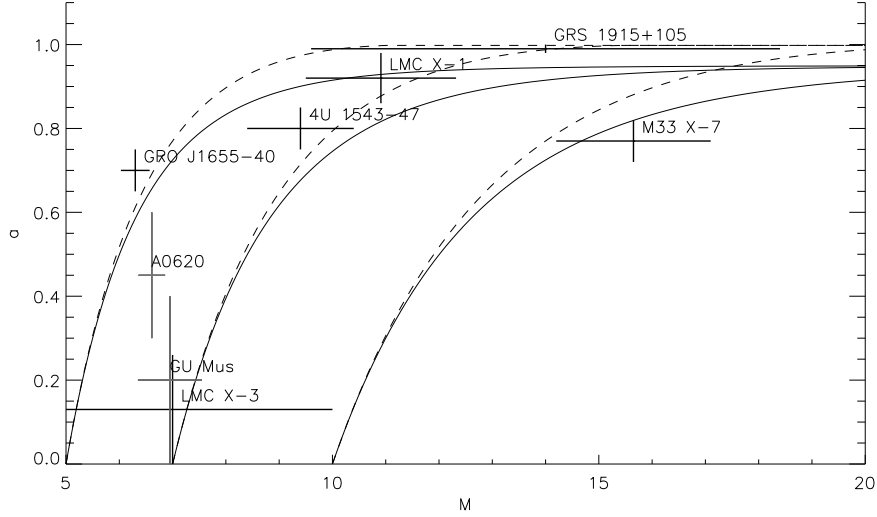}
\caption{Spin evolution for stellar mass black holes with initial masses $M_0=5M_\odot$, $M_0=7M_\odot$ and $M_0=10M_\odot$. Observation data are plotted with error bars (see \citet{Podsiadlowski}).}\label{fig:small}
\end{figure}

There are at least two methods used for black hole spin estimates: fitting the
thermal X-ray continuum and modelling the FeK$\alpha$ line profile. Due to
some effects such as comptonization the line profile can broaden and Kerr
parameter value will be overestimated. Continuum fitting, on the other hand, allows to reproduce the jet luminosity proportionality to $a^2$ \citep{narayan13}. Therefore we used fitting results \citep{Gou2010} for six objects analyzed with a continuum-fitting technique. These objects are shown in Fig.~\ref{fig:small}.

This figure can be used to constrain the possible initial masses for each object and mass gains during mass exchange, these estimates given in Table 1. Electromagnetic processes do not change initial black hole mass estimates significantly, but observational spin values may be used to constrain the contribution of electromagnetic processes and possibly the mass accretion rates.

\begin{table}\centering
\caption{ The main parameters of X-ray binary systems. Donor masses $M_d$, black hole masses $M_{BH}$, Kerr parameter $a$, estimated black hole initial masses $M^{*}_0$ and mass gains of black holes $\Delta M^{*}_{BH}$.
Black holes and donor parameters were taken from:
[0] - \citet{Gou2010}, [1]- \citet{steeghs13}, [2] - \citet{Harlaftis2004}, [3] - \citet{Shahbaz2003}, [4] - \citet{Zi2011}, [5] - \citet{Orosz2009}, [6] - \citet{Hutchings}, [7] - \citet{Orosz1998}, [8] - \citet{Valsecchi2010}, [11] - \citet{Casares1997}, [12] -\citet{xraynovae} 
}
\label{tab:time}
\bigskip
\begin{tabular}{|l|c|c|c|c|c|}
\hline
object names &  $M_{d}/M_{\odot}$ & $M_{BH}/M_{\odot}$ & $a$ & $M^{*}_{0}/M_{\odot}$& $\Delta M^{*}_{BH}/M_{\odot}$\\
\hline
LMC X-3 &$>13^{[6]}$ &    $5-11^{[0]}$  &     $<0.26^{[0]}$      &  $4.0-9.5$    &  $0.3-0.5$  \\
\hline
GRO J1655-40 &$2.36-2.94^{[3]}$ &   $6.30 \pm 0.27^{[0]}$     &      $0.65-0.75^{[0]}$    &     $4.5-5.0$  & $\sim 1.5$ \\
\hline
4U 1543-47 &$2.3-2.6^{[7]}$ &   $9.4 \pm 1.0^{[0]}$     &     $0.75-0.85^{[0]}$      &    $7.0-\sim 8.0$  & $\sim 1.3$ \\
\hline
LMC X-1 &$35-40^{[4]}$, $31.79 \pm 3.48^{[5]}$ &  $10.91 \pm 1.41^{[0]}$     &  $0.92\pm 0.06^{[0]}$        &      $6.0-7.5$ & $\sim 3.5$ \\
\hline
GRS 1915+105 &$0.47 \pm 0.27^{[1]}$, $0.81 \pm 0.53^{[2]}$ &   $14 \pm 4.4^{[0]}$     &     $0.98-1^{[0]}$    &    $5.0-9.0$ &  $5-10$  \\
\hline
M33 X-7 &$\sim 54^{[8]}$ &   $15.65\pm 1.45$    &   $0.77\pm 0.05^{[0]}$       &     $10-12^{[0]}$  &  $\sim 4$\\
\hline
GU Mus&$0.86\pm 0.075^{[9]}$ &   $6.95\pm 0.6^{[9]}$    &   $\leq 0.4^{[12]}$       &     $\sim 6-7$  &  $0.3-0.5$\\
\hline
A0620  &$0.40\pm 0.05^{[11]}$ &   $6.61\pm 0.25^{[10]}$    &   $0.3-0.6^{[12]}$       &     $5.4-5.8$  &  $\sim 1$\\
\end{tabular}
\end{table}

Results of our simulations are in accordance with the sources listed in a Table 1 except for the object GRS 1915+105. Mass gain of the black hole in this system is about 5 solar masses, but donor mass is much smaller. The black hole has a large Kerr parameter and mass, that suggests that this system could undergo component exchange  with another system. Our estimates do not take into account the possibility of such exchange that can still be important for binary evolution. The other peculiar thing about this system is the extremely high value of $a\gtrsim 0.98$ that either requires super-critical accretion or excludes the contribution of Blandford-Znajek processes during the black hole spin-up. Since this system is currently an active microquasar, the supercritical stage scenario seems more probable. 

While for microquasars, estimated spin values are usually $\gtrsim 0.5$, X-ray novae such as A~0620-00 and GRS~1124-68 are supposed to have small $a\lesssim 0.3$ \citep{xraynovae}. 
The only exception is GRO~J1655-40 that is classified both as an X-ray nova and a microquasar. Donor masses in such systems are $\lesssim 1\Msun$ and the black hole mass gains are probably insufficient for significant spin-up during the evolutionary time. 

Massive black hole binaries like M33~X-7 and the two objects from LMC have a broad range of spin values that conforms with the expectation that these objects should follow a nearly Bardeen spin-up track and thus can acquire $a \simeq 1$ if given sufficient time. 

{ Our results are in a good agreement with the recent results by \citet{Fragos14} who use models of stellar evolution to explain the observed correlation between black hole masses and rotation parameters in low-mass X-ray binary systems. }

\subsection{Monte-Carlo simulations of a SMBH population}\label{sec:moncar}

All the evolutionary scenarios described in the previous sections are simplified models. In real life black hole evolution is more complicated. We have made population synthesis of black holes rotational evolution assuming that Blandford-Znajek and DML processes may affect black hole spin evolution simultaneously. We use the Monte-Carlo technique for population synthesis of rotational evolution of a reasonably realistic supermassive black hole population.  We assume the initial masses distributed according to the log-uniform law ($dN/d\ln M = const$) between $10^6 M_{\odot}$ and $5\times 10^8 M_{\odot}$. All the objects are born at $z=15$. Dimensional accretion rate was chosen for each black hole and for each time bin according to a log-normal law with the dispersion of $ D(\lg \dot{M})=0.65$ and the mode {(probability maximum) at} $1.3\times 10^{-3}\Msunyr$. 

For these parameter values, a conventional black hole with $M \sim 10^7 M_\odot$ spends $\sim 5-10\%$ of all time in active state that conforms to the existing duty cycle estimates \citep{shankar13}.  
Initial Kerr parameters are distributed uniformly between 0 and 1. 

Here we consider the black hole active if the dimensionless accretion rate
exceeds $0.01$. This value is also assumed the transition mass accretion rate
between standard and advective disc regimes. Supercritical disc model is
applied when $\mdot > 1/\eta(a)$, where $\eta = 1-E^\dagger$ is radiative
efficiency. 

 We assume that Blandford-Znajek process operates in the circumpolar regions and DML operates near the equatorial plane of the black hole horizon.
Simulations \citep{Tchekhovskoy12,barkov12} show  that magnetic lines connecting a black hole with infinity occupy much larger surface area of the black hole horizon (Blandford-Znajek process works here). Therefore we assume that the DML works in a narrow equatorial region that occupies only $10\%$ of the black hole horizon and Blandford-Znajek process works on the remaining $90\%$ of the surface. Existence of such a region corotating with the inner part of the disc is also confirmed by simulations \citep{Narayan2013}.

\begin{figure}\label{fig:aevolv}
\centering
\includegraphics[width=0.7\columnwidth]{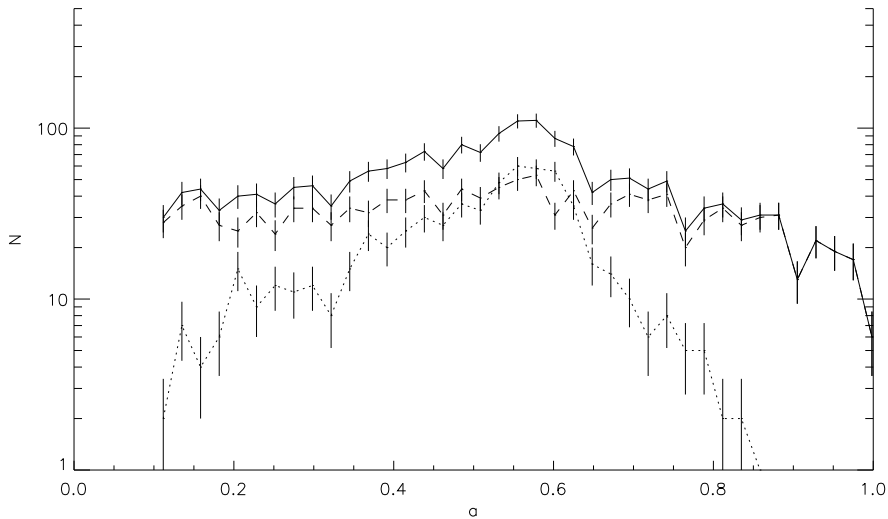}
\caption{Kerr parameter distribution for evolved supermassive black hole population. Distribution for all population is plotted by solid line, black hole in active nuclei are plotted by dashed line and in nonactive nuclei are plotted by dotted line. }\label{fig:dist1}
\end{figure}

Distribution of all the objects in Kerr parameter at $z=0$ is shown in Fig.~\ref{fig:dist1} with a solid line. Kerr parameter distribution for AGNs is plotted by a dashed line and quiescent nuclei are shown by a dotted line. 
In this figure one can see a wide distribution with a peak at  intermediate Kerr parameters $a \sim 0.6-0.7$ that corresponds to accretion through a standard disc. Supercritical accretion regime creates the high Kerr parameter wing of the distribution.  The trail to the left consists of unevolved black holes with sub-equilibrium rotation parameters of $a\sim 0.1-0.5$ and is produced by ADAF regime in quiescent BH like Sgr A*.

\begin{figure}\label{fig:distr}
\centering
\includegraphics[width=0.7\columnwidth]{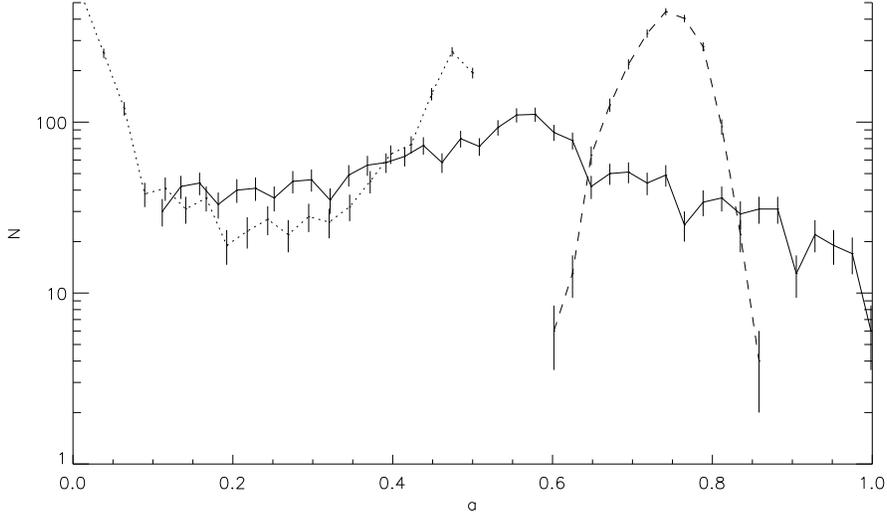}
\caption{Distribution in Kerr parameter. Distribution with initial Kerr parameter $a_0=0$ is plotted with a dotted line. Distribution with uniform initial Kerr parameter $0<a<1$ is plotted with a solid line and with a dashed line  with a large fraction ($\sim 10\%$) of supercritical regime, $a_0=0$. }\label{fig:dist2}
\end{figure}

We have also performed simulations in other assumptions, but the results
presented above seem the most realistic. In particular, we simulated black
hole population evolution affected by the BZ process only. In this case we
have only one narrow peak at $a\sim 0.5$ in contradiction with observational
data \citep{reynolds13} . Another model was evolution driven solely by the DML
process. This case is more interesting: there is a lot of objects with small
and intermediate Kerr parameters, but there are no objects with  Kerr
parameters $\gtrsim 0.7$ that disagrees with the observational data. 

In the recent work of \citet{moreno11} the author shows that in X-ray binaries
initial black hole Kerr parameters should be very small $a \sim 0$ and to
explain the observed large Kerr parameters, black holes must have suffered
long periods of  supercritical accretion. Distribution of all objects with
initial Kerr parameter uniformly distributed ($a_0<1$) is shown in
Fig.~\ref{fig:dist2}. Simulations with $a_0=0$ and a large fraction ($\sim
10\%$) of supercritical regime are plotted by a dashed line. In the same figure
distribution with uniformly distributed initial  Kerr parameter described
above is shown by a solid line. Distribution with $a_0=0$ and with a
reasonable fraction of supercritical regime ($\sim 5 \%$) is plotted by a
dotted line. This figure shows the following features for the model with $a_0
\sim 0$: black hole evolution in sense of $a(M)$ is determined by standard disc
accretion, but its timescales are longer and closer to the timescales of
inefficient ADAF accretion. Hence, a lot of objects in the simulated population are
unevolved. This model shows deficit in black holes with $a\gtrsim 0.6$ and
this is in contradiction with observations (see \citet{mcclintock11} for
review). The contribution of supercritical accretion is too small to
significantly spin up a black hole, since this regime is difficult to sustain
for black holes with $m_0>5\times 10^7$ and the fraction of supercritical
accretors with $m_0\sim 10^7$  is less than several percent. Subsequently all
the objects accumulate mass and dimensionless accetion rate
decreases. Supercritical accretion shifts the mode of the distribution towards
higher Kerr parameters. 
The model with increased mass accretion rate (shown by a dashed line in Fig.~\ref{fig:dist2})
shows a lot of rapidly rotating black holes, but there is a lack at medium and
small $a$. These two simulations are in contradiction with observations (see
\citet{reynolds13} and references therein).

\section{Discussion}\label{sec:discussion}

{ It is hard to satisfy boundary conditions in the framework of the \citet{NY95} self-similar ADAF model. Accretion disk  structure is strongly deformed near the last stable orbit where gas pressure approaches zero. In our calculations we assumed magnetic pressure proportional to thermal pressure. But while gas pressure goes to zero near the ISCO, magnetic stresses vary smoothly and remain reasonably high due to magnetic flux conservation (see \citet{Abolmasov14}). Using the self-similar model is better justified in this case than using slim-disk models together with the assumption that magnetic fields are everywhere proportional to thermal pressure. A self-consistent MHD approach would be a better approximation. 

While corrections to the electromagnetic term due to transonic nature of the disk are probably slight, angular momentum of the falling matter should be more sensitive to the processes at the inner boundary of the disk because the balance between the individual terms in the radial Euler equation shifts rapidly close to the sonic surface. 

The matter falling inside the last stable orbit conserves its angular momentum while outside the sonic surface, angular momentum may be easily transferred by tangential viscous stresses. Hence the angular momentum of the in-falling matter should be estimated at the sonic surface rather than at the ISCO. 

Continuity equation may be written approximately as:

\begin{equation}
\displaystyle\frac{d(r^d \rho v_r)}{dr}=0,
\end{equation}

where $d$ describes the geometry of the streamlines: $d=1$ corresponds to cylindrical radial  flow, $d=2$ in case of spherical radial flow.

Let us combine it with the radial component of the Euler equation:

\begin{equation}
v_r \displaystyle\frac{d v_r}{dr}-\displaystyle\frac{v^2_{\phi}}{r} = -\displaystyle\frac{1}{\rho}\displaystyle\frac{dp}{dr}-\displaystyle\frac{v^2_K}{r}
\end{equation}

Using these equations one can obtain the radial Grad-Shafranov equation \citep{Beskinbook}:

\begin{equation}
\displaystyle\frac{d \ln \rho}{d \ln r}= \displaystyle\frac{v^2_{\phi}- v^2_K+d\cdot v^2_r}{r c^2_s(1-\displaystyle\frac{v^2_r}{c^2_s})}
\end{equation}

At the sonic surface, denominator becomes zero hence the numerator should be zero too. This yields us angular momentum in the form:

\begin{equation}
L=L_K\sqrt{1-\displaystyle\frac{d v^2_r r}{GM}}=L_K\sqrt{1-\displaystyle\frac{d c^2_s r}{GM}}=L_K\sqrt{1-d \cdot \displaystyle\left(\frac{H}{r}\right)^2}=c^{*}_1 L_K
\end{equation}

Here $c^{*}_1$ is coefficient that describes the deviations from Keplerian angular law (see section~3.3). 
In the self-similar ADAF model, $(H/r)^2=2.5c_3$. In our calculations we used $c_3=0.285$. Using this value for cylindrical radial flow one can find $c^{*}_1=0.536$ that is close to the value $c_1=0.43$ that we used in our calculations. For spherical flow $c^{*}_1$ is very close to unity.  
We may conclude that our calculations are close to reality if disk height is constant near the sonic surface.  

This may be the case if the flow inside the last stable orbit becomes magnetically supported in vertical direction.

In \citet{Shafee08}, \citet{Abolmasov14} and other works, rotation of magnetized accretion disks was found to be close to Keplerian near the last stable orbit. However, these works aimed on reproducing relatively thin disks with $H/r \sim 0.1$. In thicker disks, deviations from Keplerian law should be larger. Besides, magnetic field configuration in fact is expected to affect the rotation of the inner disk regions through magnetic pressure and tangential stress. }

In general, recent simulations support the BZ scenario \citep{Tchekhovskoy12} and practically exclude the DML scenario. There are two reasons for this. First, inside the last stable orbit, density becomes very small and thermal pressure can no more counteract the Maxwellian stresses that leads to deformation and expansion of the magnetic field lines toward infinity.
The second reason is existence of a large magnetic field in Blandford-Znajek configuration (with geometry close to a split-monopole) that leaves very small area for accretion and for field lines connecting the stretched horizon with the accretion disc. Numerical results suggest that either DML contribution to black hole rotation is practically absent or at least operates only on a small fraction of the stretched horizon surface (see also above section ~\ref{sec:moncar}).

However, the results of these simulations are in apparent contradiction with observational data on super-massive black holes in galactic nuclei. 
Observations show that only about $7\div 8\%$ of active galactic nuclei launch radio-bright jets \citep{Ivezi02}. The reason for prevalence of radio-quiet objects is yet unclear, it may be connected to the overall magnetic flux through the black hole vicinity or to the broken symmetry of the magnetic field geometry. 
A probable reason for the discrepancy between simulations and observations is that most of the recent numerical simulations consider accretion discs with large-scale poloidal magnetic fields capable for rapid accumulation of magnetic flux inside the last stable orbit. Perhaps, smaller-scale chaotic magnetic fields would produce field geometry that is closer to the DML case. 

\bigskip

{ 
DML configuration plotted in Fig. 1 can not be realized because Lorentz force acting on the magnetic loops inside the free-fall region will collapse them vertically. 
A more realistic picture with both processes, BZ and DML, taking place, is plotted in Fig. 11. Magnetic fields are supposed to be chaotic in the accretion disc: even regular fields are converted to small-scale loops due to Hall effect and magnetic instabilities (such as the magneto-rotational instability applied for accretion discs by \citet{MRI}). Hence, in the accretion flow, substantial part of the magnetic field energy will be stored in small-scale loops having zero magnetic flux through the equatorial plane of the black hole. This part of the magnetic field is also a subject to dissipation at the horizon. As long as these chaotic magnetic fields are confined inside the accretion flow they can contribute to the DML process. { Note that any magnetic loop with non-zero radial extension contributes to DML through frame-dragging; in full general relativity, one does not need to approach the horizon.} Whether this condition is normally fulfilled or not is non-trivial and requires further thorough analytical and numerical consideration. 
}

\begin{figure}\label{fig:9}
\centering
\includegraphics[width=0.7\columnwidth]{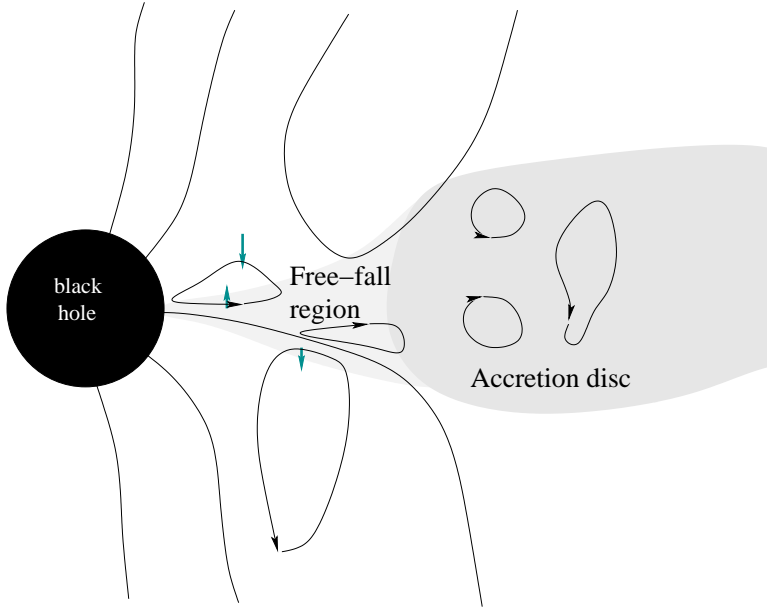}
\caption{More realistic magnetic field configuration.  Lorentz forces are shown as blue (grey) arrows.                                 } 
\end{figure}

In our paper we did not consider mergers that may however have an important effect upon the evolution of SMBH. For example, in the work of \citet{volonteri05} the authors have made population synthesis of black holes with mergers and obtained Kerr parameter distribution similar to our result: Kerr parameter distributed between 0.1 and 0.9 with a peak at $a=0.5-0.6$. Most probably, if we will perform population synthesis of a black hole population taking into consideration both mergers and electromagnetic processes, we would obtain a qualitatively similar but broader distribution. But the question remains open: why do we observe a lot of black holes with high Kerr parameter? And the most probable explanation is supercritical accretion.

For a sample of black holes accreting magnetized matter for a large enough amount of time, broad distribution in spin is expected. Some of the objects stall near the critical $a$ value, some are expected to evolve toward $a\sim 1$. However, spin and mass evolution timescales are mostly larger than the Eddington timescale, and often larger than Hubble time. Together with the importance of mergers for black holes in galactic nuclei, it makes unreasonable to expect black hole rotation to be consistent with any unique spin value. Modeling taking into account all the effects is required. 

In general, any black hole that gains an amount of mass comparable to its initial mass approaches its equilibrium rotation. Higher-mass black holes in binary systems should have also higher Kerr parameters, in consistence with the observed higher masses of the black holes in microquasars in comparison with the lighter black holes in X-ray novae \citep{banerjee13}.

\section{Conclusions}\label{sec:conclusion}

In our work we aimed on constraining the role of angular momentum exchange
between the BH and accretion disc and that of general relativistic magnetic
field decay that should affect this process. We find that DML-like processes
may be important for thick advective discs, but magnetic field is damped significantly close to the horizon.
 Equilibrium Kerr parameter in this case is in the range $\sim 0.3-0.5$ and is
 primarily determined by deviations from Keplerian law in the disc. 
In supercritical regime, both electromagnetic processes do not influence black hole spin evolution significantly.

We have made Monte-Carlo simulations for a supermassive black hole population assuming that both processes operate simultaneously. For a reasonable parameter set, we obtain a broad distribution in Kerr parameter $0.1\lesssim a \lesssim 0.9$ with a peak at $a\sim 0.7$ that is in fair agreement with observational data. 
For stellar mass black holes, the observed correlation between black hole masses and Kerr parameters may be understood as a consequence of accretion mass gain of $1-10 M_{\odot}$ in most of black hole binary systems.

Our results show that black hole observational appearance is sensitive to the accretion history, primarily to the period when most of the mass is gained. In the standard disc regime, a black hole gains a larger part of its mass and Kerr parameter increases faster than in ADAF regime.

\section*{Acknowledgments}

 Anna thanks to RSF grant 14-12-00146 and Pasha also thanks to RFBR grant 14-02-91172 and Dynasty foundation. We would like to thank Professor Nikolai Shakura for inspiring of the basic idea of this work, and Professor Vasily S. Beskin and Sasha Tchekhovskoy for fruitful discussions on black hole magnetospheres. We are also grateful to the anonymous referees for valuable comments.

\bibliographystyle{mn2e}
\bibliography{mybib}

\label{lastpage}

\end{document}